\documentclass[10pt,final,journal]{IEEEtran}
\usepackage[switch]{lineno} 
\usepackage{tabularx} 
\usepackage{booktabs}
\usepackage{multirow}
\usepackage{multicol}
\usepackage{makecell}

\usepackage{stmaryrd}
\usepackage{graphicx}
\usepackage{amsthm} 
\usepackage{amsmath} 
\usepackage{bm}
\usepackage{graphicx}
\usepackage{epstopdf}
\usepackage{color}
\usepackage{float}
\usepackage[colorlinks=false,linkcolor=black,urlcolor=black,bookmarksopen=true, hidelinks]{hyperref}
\usepackage{bookmark}
\usepackage[flushleft]{threeparttable}
\usepackage{stackengine}
\usepackage{scalerel}
\usepackage{array}

\usepackage{xcolor,soul}

\usepackage{cite}

\ifCLASSINFOpdf
\else
\fi
\usepackage{amsfonts,amssymb}

\usepackage{amsmath}
\usepackage{algorithm}
\usepackage{algpseudocode}
\ifCLASSOPTIONcompsoc
 \usepackage[caption=false,font=normalsize,labelfont=sf,textfont=sf]{subfig}
\else
 \usepackage[caption=false,font=footnotesize]{subfig}
\fi
\begin{document}
%

\title{New Closed-form Joint Localization and Synchronization using Sequential One-way TOAs}

\author{Ningyan~Guo,
		Sihao~Zhao, 
		Xiao-Ping~Zhang, \textit{Fellow, IEEE,}
		Zheng~Yao,
		Xiaowei~Cui,
        and~Mingquan~Lu
\thanks{This work was supported in part by the National Natural Science Foundation of China under Grant 62103226. \textit{(Corresponding authors: Sihao Zhao; Zheng Yao)}}
\thanks{N. Guo is with with the School of Information and Communication Engineering, Beijing University of Posts and Telecommunications, Beijing 100876, China (e-mail: guoningyan@bupt.edu.cn).}
\thanks{Z. Yao, X. Cui are with the Department of Electronic Engineering,
	Tsinghua University, Beijing 100084, China (e-mail: yaozheng@tsinghua.edu.cn; cxw2005@tsinghua.edu.cn).}
\thanks{S. Zhao, X.-P. Zhang are with the Department of Electrical, Computer and Biomedical Engineering, Ryerson University, Toronto, ON M5B 2K3, Canada (e-mail: sihao.zhao@ryerson.ca; xzhang@ryerson.ca).}
\thanks{M. Lu is with the Department of Electronic Engineering,
	Beijing National Research Center for Information Science and Technology, Tsinghua University, Beijing 100084, China. (e-mail: lumq@tsinghua.edu.cn).}
}

\markboth{}%
{Shell \MakeLowercase{\textit{et al.}}: Bare Demo of IEEEtran.cls for IEEE Journals}
%




\maketitle

\begin{abstract}
It is an essential technique for the moving
user nodes (UNs) with clock offset and clock skew to
resolve the joint localization and synchronization (JLAS)
problem.
Existing iterative maximum likelihood methods using sequential one-way time-of-arrival (TOA) measurements from the anchor nodes' (AN) broadcast signals require a good initial guess and have a computational complexity that grows with the number of iterations, given the size of the problem. In this paper, we propose a new closed-form JLAS approach, namely CFJLAS, which achieves the asymptotically optimal solution in one shot without initialization when the noise is small, and has a low computational complexity. After squaring and differencing the sequential TOA measurement equations, we devise two intermediate variables to reparameterize the non-linear problem. In this way, we convert the problem to a simpler one of solving two simultaneous quadratic equations. We then solve the equations analytically to obtain a raw closed-form JLAS estimation.
Finally, we apply a weighted least squares (WLS) step to optimize the estimation. We derive the Cram\'er-Rao lower bound (CRLB), analyze the estimation error, and show that the estimation accuracy of the CFJLAS reaches the CRLB under the small noise condition. The complexity of the new CFJLAS is only determined by the size of the problem, unlike the conventional iterative method, whose complexity is additionally multiplied by the number of iterations. Simulations in a 2D scene verify that the estimation accuracies of the new CFJLAS method in position, velocity, clock offset, and clock skew all reach the CRLB under the small noise condition. Compared with the conventional iterative method, which requires a proper initialization to converge and has a growing complexity with more iterations, the proposed new CFJLAS method does not require initialization, obtains the optimal solution under the small noise condition, and has a low computational complexity.

\end{abstract}



%
\IEEEpeerreviewmaketitle

\section{Introduction}\label{Introduction}
%
%
%
%
\IEEEPARstart{L}{ocalization} and synchronization systems are playing a fundamental role in our daily lives. In such systems, there are usually multiple anchor nodes (ANs) with known locations and user nodes (UNs) to be localized and synchronized. It is widely adopted in a variety of applications such as Internet of Things (IoT), emergency rescue, aerial surveillance and target detection and tracking \cite{kuutti2018survey,ferreira2017localization,beard2002coordinated}.
To enable the positioning, navigation and timing for UNs in the absence of the external spatiotemporal coordinate system, joint localization and synchronization (JLAS) in a relative spatiotemporal coordinate system formed by the ANs is an important technology. To achieve JLAS for the UNs, the ANs transmit signals from known places and the UNs capture the signals to obtain measurements, such as time-of-arrival (TOA) and time-of-flight (TOF). The TOF measurement requires perfect synchronization between the AN and UN, which is costly to implement. The TOA measurement is comprised of the true distance along with the clock offset between the ANs and the UN, and thus does not require synchronization between the AN and UN. It is one of the most widely adopted measurements for JLAS due to its simplicity and high accuracy \cite{denis2006joint,vaghefi2015cooperative,ahmad2013joint,etzlinger2017cooperative,luo2013new,zheng2010joint}.

Code division and frequency division are among the most widely used TOA schemes. Two well-known examples of these two schemes are the Global Positioning System (GPS), which adopts code division scheme, and the GLONASS, which uses frequency division \cite{hofmann2007gnss,misra2006global}. However, code division scheme suffers from the near-far effect especially in a small region where a UN is near the AN \cite{picois2014near}, and frequency division scheme needs to separate the signals from ANs into different frequency bands \cite{myung2006single} and increases the complexity of the radio frequency front-end design, especially for a small and low-cost UN.

Unlike code division or frequency division schemes, a time division (TD) scheme generating sequential TOA measurements is free of the near-far effect or large bandwidth occupancy, although it would require a longer temporal duration for the ANs to finish their signal transmission. A UN in a TD broadcast system obtains the sequential one-way TOA measurements to achieve JLAS on its own, without any transmission back to the ANs. Such a TD broadcast scheme reduces the reception channels of a UN, supports an unlimited number of UNs, and offers higher safety for UNs from being detected. Hence, an increasing number of research works on the TD scheme have appeared  \cite{dwivedi2012scheduled,shi2019blas,dwivedi2013cooperative,hamer2018self,zachariah2013self,carroll2014demand,yan2017asynchronous,zhao2020optloc,shi2020sequential}.

Localization based on the TD broadcast sequential TOA has been studied in recent years. Dwivedi et al. \cite{dwivedi2013cooperative} develop a solution for cooperative localization by using scheduled wireless transmissions. They formulate two practical estimators, namely sequential weighted least squares multi-dimensional scaling (WLS-MDS) and maximum likelihood (ML), for localization and derive the Cram\'er-Rao lower bound (CRLB) to evaluate the performance of these two estimators. Zachariah et al. \cite{zachariah2013self} propose a self-localization method for a receiver node using given sequential measurements based on an iterative maximum a posterior (MAP) estimator that considers the uncertainty of the ANs and the turn-around delay of the asynchronous transceivers. They also propose a sequential approximate maximum a posterior estimator to resolve the localization problem based on sequential TOAs \cite{zachariah2014schedule}. Carroll et al. \cite{carroll2014demand} use a sequential transmission protocol to resolve an on-demand asynchronous localization problem for an underwater acoustic network, and verify through simulation that passive nodes can achieve low-error positioning. Yan et al. \cite{yan2017asynchronous} investigate a localization method for passive nodes based on sequential messages. Simulation results show that the proposed iterative least squares (LS) method can improve the localization accuracy compared with the synchronous localization algorithm. However, all these methods only resolve the node's localization problem and do not address the clock synchronization issue.

For JLAS using the sequential one-way TOA measurements, several iterative methods are proposed. Shi et al. \cite{shi2019blas} present a distributed state estimation method for the UNs, which utilizes the TD broadcast inter-node information to resolve the JLAS problem. Their method enables accurate, high-frequency and real-time estimation, and supports an unlimited number of UNs. However, their iterative method has a high computational complexity and does not apply to a moving UN. Shi et al. \cite{shi2020sequential} extend the previous work of \cite{shi2019blas}, and utilize a two-step weighted least squares (WLS) method to jointly estimate the position, velocity and clock parameters of the UNs in the presence of the position uncertainties of ANs. 
However, this method requires to transform the localization problem into a higher-dimensional state space, and thus needs more measurements and has a high computational complexity due to the second iterative step. Zhao et al. \cite{zhao2020optloc} develop an ML method and an iterative algorithm to estimate the localization and synchronization parameters. However, this method requires a good initial guess and has a high computational cost due to iteration.

Unlike the iterative methods mentioned above, the closed-form JLAS estimators require no initial guess or iteration, and have low computational complexity. There are extensive studies on closed-form methods for concurrent TOA systems. Bancroft originally introduces a closed-form approach in solving the GPS equations \cite{bancroft1985algebraic}. Compared with the iterative methods, Bancroft's closed-form algorithm significantly reduces the computational complexity without initialization. A few extensions of Bancroft's algorithm to the case with noisy time-difference-of-arrival (TDOA) measurements are proposed in \cite{smith1987closed,schau1987passive,smith1987spherical}. Chan and Ho \cite{chan1994simple} propose a two-step WLS (TSWLS) estimator that achieves the CRLB at a small noise level. Closed-form localization methods based on the multidimensional scaling technique that utilizes a squared distance matrix are proposed in \cite{cheung2005multidimensional,so2007generalized}. Zhu and Ding \cite{zhu2010joint} generalize Bancroft's algorithm to the case with more sensors and noisy measurements. Other efficient closed-form methods are developed in \cite{huang2013efficient,sun2013joint} for joint localization and synchronization tasks. Wang et al. \cite{wang2015toa} use a closed-form method to resolve the JLAS problem when the known sensor positions and clock biases are subject to random errors. Zhao et al. \cite{zhao2020closed} propose a closed-form localization method using TOA measurements from two non-synchronized systems. However, these closed-form methods all assume that the TOA measurements are concurrent and thus are not applicable to the system with sequential one-way TOA messages.

In this paper, we propose a new closed-form method, namely CFJLAS, to resolve the JLAS problem for the moving UNs with clock offset and skew using sequential one-way TOA measurements in a wireless broadcast JLAS system. We first transform the nonlinear relation between the measurements and the parameters to be estimated by squaring and differencing the sequential TOA measurement equations. We then devise two intermediate variables to reparameterize the relation, and construct two simultaneous quadratic equations with respect to the intermediate variables. Then, we find the solution of the intermediate variables by solving the quadratic equations analytically. We then apply the LS method to roughly estimate positions and the clock parameters of the moving UNs from the results of the intermediate variables. Finally, we apply a WLS step using the residuals of all the measurements to obtain the final optimal estimation. We derive the CRLB of the estimation error for the moving UN and show that the estimated results from the new CFJLAS reach the CRLB when the measurement noise is small. Simulation of a 2D moving scene is conducted to verify the performance of the new CFJLAS. Numerical results show that the accuracies for the position, velocity, clock offset and clock skew from the proposed CFJLAS all reach the CRLB. Compared with the iterative ML method (LSPM-UVD) \cite{zhao2020optloc}, which requires accurate initialization and has an increasing complexity with the growing number of iterations, the new method obtains the optimal solution in one shot without requiring an initial guess under the small noise condition. Given the number of ANs, the new CFJLAS method has a low computational complexity, which is at the same level as three iterations of the conventional iterative ML method. 

The rest of the paper is organized as follows. In Section II, the sequential TOA-based JLAS problem model for the moving UNs is formulated. A new closed-form JLAS method is proposed in detail in Section III. Then, we derive the CRLB, analyze the theoretical error and complexity of the new method in Section IV. Simulations are conducted to evaluate the performance of this new CFJLAS method in Section V. Finally, Section VI concludes the paper.

The main notations used in this paper are summarized in Table \ref{table_notation}.

\begin{table}[!t]
	\caption{Notation List}
	\label{table_notation}
	\centering
	\begin{tabular}{l p{5.5cm}}
		\toprule
		lowercase $x$&  scalar\\
		bold lowercase $\boldsymbol{x}$ & vector\\
		bold uppercase $\bm{X}$ & matrix\\
		$\Vert \boldsymbol{x} \Vert$ & Euclidean norm of a vector\\
		$i$, $j$ & indices of variables\\
		$[\boldsymbol{x}]_{i}$ &the $i$-th element of a vector\\
		$[\bm{X}]_{i,:}$, $[\bm{X}]_{:,j}$ &the $i$-th row and the $j$-th column of a matrix, respectively\\
		$[\bm{X}]_{i,j}$ &entry at the $i$-th row and the $j$-th column of a matrix\\
		$|\bm{X}|$ &determinant of a matrix\\
		$\mathbb{E}[\cdot]$ & expectation operator \\
		$\mathrm{diag}(\cdot)$ & diagonal matrix with the elements inside\\
		$M$ & number of ANs\\
		$K$ & dimension of all the position and velocity vectors, i.e., $K=2$ in 2D case and $K=3$ in 3D case\\
		$\boldsymbol{p}_{i}$, $\beta_i$& position vector and clock offset of AN \#$i$\\
		$\boldsymbol{p}$, $\boldsymbol{v}$ &  unknown position and velocity vector of a UN\\
		$\beta$, $\omega$ & unknown clock offset and skew of a UN\\
		$\boldsymbol{\theta}$ &  parameter vector, $\boldsymbol{\theta}=[\boldsymbol{p}^T,\boldsymbol{v}^T,\beta,\omega]^T$\\
		$\lambda_1$, $\lambda_2$ & intermediate variables\\
		$\hat{\tau}_{i}$ & noisy sequential TOA measurement with respect to AN \#$i$\\
		$\hat{\boldsymbol{\tau}}$ & collective form of TOA measurement $\hat{\tau}_i$\\
		$t_i$ & time interval from the beginning of the TD broadcast round to AN \#$i$'s transmission time\\
		$\varepsilon_i$, $\sigma_i^2$ &  Gaussian random noise and variance for the $i$-th TOA measurement\\
		$\boldsymbol{\varepsilon}$ & collective form of noise term $\varepsilon$\\
		$\boldsymbol{l}_i$ & unit line-of-sight (LOS) vector from the UN to AN \#$i$\\
		$\bm{\mathcal{F}}$ &  Fisher information matrix\\
		$\bm{C}$ & inverse of the covariance matrix of noise $\boldsymbol{\varepsilon}$\\
		$\bm{W}$ & weighting matrix\\
		$\bm{J}$ & design matrix\\
		$\bm{O}_{M\times N}$ & $M\times N$ matrix with all entries being zero\\
		$\bm{I}_{M}$ & $M\times M$ identity matrix\\
		\bottomrule
	\end{tabular}
\end{table}

\section{Problem Statement}
\begin{figure}
	\centering
	\includegraphics[width=1\linewidth]{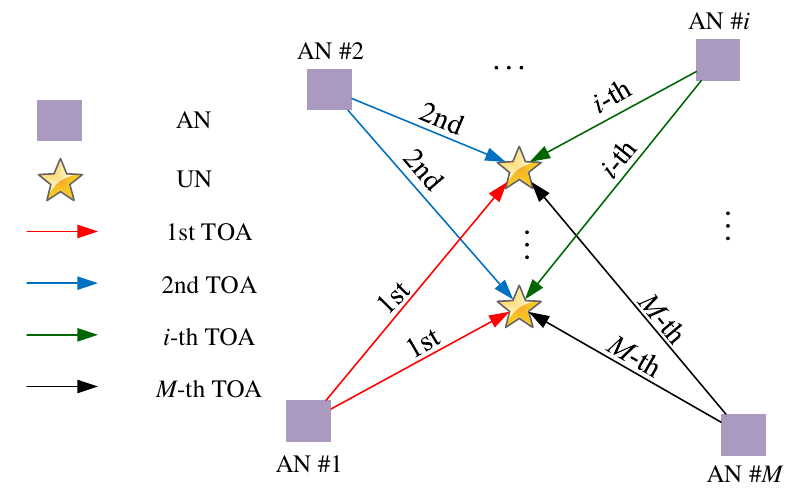}
	\caption{Joint localization and synchronization (JLAS) system using the TD broadcast scheme. The system is comprised of ANs and UNs. ANs act as moving beacons by broadcasting on-air packets sequentially in scheduled time-slots. UNs only passively receive the on-air packets to obtain the sequential one-way TOA measurements, and then achieve localization and synchronization.
	}
	\label{fig:sysfig}
\end{figure}

We consider a system comprised of $M$ ANs and a number of moving UNs, as depicted in Fig. \ref{fig:sysfig}. It operates in a wireless broadcast manner using a TD scheme. All ANs are synchronized to a system clock. They sequentially broadcast the on-air packets according to its pre-scheduled launch time slots, e.g., AN \#1, AN \#2 and so on. The ANs set up a relative spatiotemporal coordinate system in advance. The UNs receive the on-air packets to obtain the sequential TOA measurements and achieve JLAS in this relative spatiotemporal coordinate system. Therefore, we aim to resolve the moving UNs' JLAS problem based on the observed sequential TOA measurements in a wireless broadcast network.

Without loss of generality, we take one moving UN as an example to present the JLAS problem. Other moving UNs in the system work in the same manner. We investigate the case with line-of-sight (LOS) signal propagation between the ANs and the moving UN. The sequential TOA measurements in one TD broadcast round from all $M$ ANs are used. Additionally, since the communication rate or bandwidth is high enough, one TD broadcast round for all ANs is short, usually at the millisecond level. We model that the moving UN's velocity remains constant in the period of one TD broadcast round.

We denote the unknown position, velocity, clock offset and clock skew of a UN at the beginning of one TD broadcast round by $\boldsymbol{p}$, $\boldsymbol{v}$, ${\beta}$, and $\omega$, respectively. The true position and clock offset of AN \#$i$, $i=1,\cdots,M$, are denoted by $\boldsymbol{p}_i$ and ${\beta_i}$, respectively. In a generic case, the position of AN \#$i$, denoted by $\hat{\boldsymbol{p}}_i$ is usually measured or estimated, and thus may contain error, denoted by $\delta \bm{p}_i={\boldsymbol{p}}_i-\hat{\boldsymbol{p}}_i$. We assume that the error follows a Gaussian distribution, i.e., $\delta \boldsymbol{p}_i \sim \mathcal{N}(\bm{0},\bm{\Sigma}_{i})$.
The position and velocity are both of $K$ dimensional, $K=2$ or $K=3$. Given that the velocity of the UN is constant during one TD broadcast round, the one-way TOA measurement from AN \#$i$ is expressed by
\begin{align}\label{eq:TOA}
	\hat{\tau}_i&=\left\Vert\boldsymbol{p}+\boldsymbol{v}t_i-\boldsymbol{p}_i\right\Vert+\beta+\omega t_i-\beta_i+\varepsilon_i\nonumber\\
	&=\left\Vert\boldsymbol{p}+\boldsymbol{v}t_i-\hat{\boldsymbol{p}}_i-\delta \bm{p}_i\right\Vert+\beta+\omega t_i-{\beta}_i+\varepsilon_i, \nonumber\\
	&i=1,\cdots,M,
\end{align}
where $\hat{\tau}_i$, $\beta$, $\beta_i$ and $\varepsilon_i$ have the unit of meter, $\omega$ has the unit of meter per second, $t_i$ is the interval between the start time of the TD broadcast round and AN \#$i$'s launch time, and $\varepsilon_i$ is the measurement noise, following an independent zero-mean Gaussian distribution with a variance of $\sigma_i^2$, i.e., $\varepsilon_i \sim \mathcal{N}(0,\sigma_i^2)$.


We aim to find an accurate estimation of the moving UN's position and clock parameters using all the sequential TOA measurements given by (\ref{eq:TOA}). It is written as
\begin{align}
	\text{Given } \{\hat{\tau}_i,\hat{\bm{p}}_i\}_{i=1}^M, \text{ estimate } \{\bm{p},\bm{v},\beta,\omega\}.
\end{align}

To solve this problem, the parameters $\boldsymbol{p}$, $\boldsymbol{v}$, ${\beta}$, and $\omega$ need to be estimated. The existing iterative ML method in \cite{zhao2020optloc} requires a proper initial guess to obtain accurate estimation and has a computational complexity, which grows with the increasing number of iterations. We present a new closed-form method, which does not require initialization or iteration and has a lower complexity than the iterative method, to resolve this problem in the next section.

\section{A New Closed-form JLAS Method}
In this section, we develop a new closed-form JLAS method, namely CFJLAS. Firstly, we reparameterize the nonlinear problem. By employing two intermediate variables, we transform the problem to a simpler one of solving a pair of quadratic equations. Then, we find the solution of the intermediate variables, based on which a raw estimation of the moving UN's position and clock parameter is obtained. Finally, we apply a WLS step to obtain the final optimal estimation. The above steps are presented in detail in the following sub-sections.

\subsection{Step 1: Reparameterization}

To obtain a closed-form solution, we first re-organize and square the two sides of (\ref{eq:TOA}), and come to
\begin{align}\label{eq:tausquare}
\left(\hat{\tau}_i+\beta_i-\beta-\omega t_i-\varepsilon_i\right)^2=\left\Vert\boldsymbol{p}+\boldsymbol{v}t_i-\hat{\boldsymbol{p}}_i-\delta\bm{p}_i \right\Vert^2
\end{align}

We define $\hat{\alpha}_i\triangleq\hat{\tau}_{i}+\beta_i$, expand and re-organize (\ref{eq:tausquare}), and come to
\begin{align}\label{eq:tausquare1}
&2\hat{\bm{p}}_i^T\boldsymbol{p}+2t_i\hat{\bm{p}}_i^T\boldsymbol{v}-2\hat{\alpha}_i\beta-2t_i\hat{\alpha}_i\omega+\left(\beta^2-\Vert\boldsymbol{p}\Vert^2\right)\nonumber\\
=&\Vert\hat{\bm{p}}_i\Vert^2-\hat{\alpha}_i^2-t_i^2\left(\omega^2-\Vert\boldsymbol{v}\Vert^2\right)-2t_i\left(\beta\omega-\boldsymbol{p}^T\boldsymbol{v}\right)+{\eta}_i+\varepsilon_{i}^2,\nonumber\\
&i=1,\cdots,M,
\end{align}
where ${\eta}_i=2\left(\hat{\alpha}_i-\beta-\omega t_i\right)\varepsilon_i-2\left(\bm{p}^T+t_i\bm{v}^T-\hat{\bm{p}}_i^T\right)\delta \bm{p}_i$.

To remove the term $\beta^2-\Vert\boldsymbol{p}\Vert^2$, we can conduct subtraction between equation (\ref{eq:tausquare1}) with arbitrary $i$ and equations (\ref{eq:tausquare1}) with other $i$ values. Without loss of generality, we substitute $i=1$ into (\ref{eq:tausquare1}) and have 
\begin{align}\label{eq:tausquare2}
	&2\hat{\bm{p}}_1^T\bm{p}+2t_1\hat{\bm{p}}_1^T\boldsymbol{v}-2\hat{\alpha}_1\beta-2t_1\hat{\alpha}_1\omega+\left(\beta^2-\Vert\boldsymbol{p}\Vert^2\right)=\nonumber\\
	&\Vert\hat{\bm{p}}_1\Vert^2-\hat{\alpha}_1^2-t_1^2\left(\omega^2-\Vert\boldsymbol{v}\Vert^2\right)-2t_1\left(\beta\omega-\boldsymbol{p}^T\boldsymbol{v}\right)+{\eta}_1+\varepsilon_1^2.
\end{align}

By subtracting (\ref{eq:tausquare2}) from (\ref{eq:tausquare1}), we remove the term $\beta^2-\Vert\boldsymbol{p}\Vert^2$. We put all the linear unknown parameters on the left and have
\begin{align}\label{eq:difftausquare}
&2\left(\hat{\bm{p}}_i^T-\hat{\bm{p}}_1^T\right)\boldsymbol{p}+2\left(t_i\boldsymbol{p}_i^T-t_1\boldsymbol{p}_1^T\right)\boldsymbol{v}+2\left(\hat{\alpha}_1-\hat{\alpha}_i\right)\beta+ \nonumber\\
&2\left(t_1\hat{\alpha}_1-t_i\hat{\alpha}_i\right)\omega\nonumber\\
=&\Vert\hat{\bm{p}}_i\Vert^2-\Vert\hat{\bm{p}}_1\Vert^2-\left(\hat{\alpha}_i^2-\hat{\alpha}_1^2\right)+\left(t_1^2-t_i^2\right)\left(\omega^2-\Vert\boldsymbol{v}\Vert^2\right)+\nonumber\\
& 2\left(t_1-t_i\right)\left(\beta\omega-\boldsymbol{p}^T\boldsymbol{v}\right)+{\eta}_i-{\eta}_1+\varepsilon_i^2-\varepsilon_1^2,\nonumber\\
&i=2,\cdots,M.
\end{align}

There are still unknown terms including $\omega^2-\Vert\boldsymbol{v}\Vert^2$ and $\beta\omega-\boldsymbol{p}^T\boldsymbol{v}$ in \eqref{eq:difftausquare}. It is not possible to employ only one extra variable to transform \eqref{eq:difftausquare} into a linear relation as was done in \cite{smith1987closed,schau1987passive,smith1987spherical}, since we have more parameters to estimate. We therefore devise two intermediate variables as
\begin{align} \label{eq:lambda}
\lambda_1&=\omega^2-\Vert\boldsymbol{v}\Vert^2,\nonumber\\
\lambda_2&=\beta\omega-\boldsymbol{p}^T\boldsymbol{v}.
\end{align}

We denote the unknown parameters by $\boldsymbol{\theta}=[\boldsymbol{p}^T,\boldsymbol{v}^T,\beta,\omega]^T$. By collecting terms, squaring and taking differences as given by \eqref{eq:difftausquare}, we transform the set of equations in \eqref{eq:TOA} into the collective form as
\begin{align}\label{eq:collective1}
\hat{\bm{A}}\boldsymbol{\theta}=\hat{\bm{y}}+\bm{G}\left[\lambda_1(\boldsymbol{\theta}),\lambda_2(\boldsymbol{\theta})\right]^T+\boldsymbol{\eta}_d(\boldsymbol{\theta})+\bm{z},
\end{align}
where
\begin{align} \label{eq:Amat}
&\hat{\bm{A}}=\nonumber\\
&2\left[
\begin{matrix}
\hat{\bm{p}}_2^T-\hat{\bm{p}}_1^T & t_2\hat{\bm{p}}_2^T-t_1\hat{\bm{p}}_1^T & \hat{\alpha}_1-\hat{\alpha}_2&t_1\hat{\alpha}_1-t_2\hat{\alpha}_2\\
\vdots &\vdots &\vdots&\vdots\\
\hat{\bm{p}}_M^T-\hat{\bm{p}}_1^T & t_M\hat{\bm{p}}_M^T-t_1\hat{\bm{p}}_1^T & \hat{\alpha}_1-\hat{\alpha}_M&t_1\hat{\alpha}_1-t_M\hat{\alpha}_M\\
\end{matrix}
\right],
\end{align}
\begin{align} \label{eq:matG}
\bm{G}=\left[
\begin{matrix}
t_1^2-t_2^2 & 2(t_1-t_2)\\
\vdots &\vdots\\
t_1^2-t_M^2 & 2(t_1-t_M)\\
\end{matrix}
\right],
\end{align}
\begin{align} \label{eq:vectory}
\hat{\bm{y}}=\left[
\begin{matrix}
\Vert\hat{\bm{p}}_2\Vert^2-\Vert\hat{\bm{p}}_1\Vert^2-\left(\hat{\alpha}_2^2-\hat{\alpha}_1^2\right)\\
\vdots\\
\Vert\hat{\bm{p}}_M\Vert^2-\Vert\hat{\bm{p}}_1\Vert^2-\left(\hat{\alpha}_M^2-\hat{\alpha}_1^2\right)
\end{matrix}
\right],
\end{align}
\begin{align}
\boldsymbol{\eta}_d=\left[
\begin{matrix}
\eta_2-\eta_1,&\cdots,&\eta_M-\eta_1
\end{matrix}
\right]^T,
\end{align}
\begin{align}
	\boldsymbol{z}=\left[
	\begin{matrix}
		\varepsilon_2^2-\varepsilon_1^2,&\cdots,&\varepsilon_M^2-\varepsilon_1^2
	\end{matrix}
	\right]^T.
\end{align}
Note that $\lambda_1$, $\lambda_2$, and $\boldsymbol{\eta}_d$ have non-linear dependence on some of the elements of $\bm{\theta}$, and thus are written as functions of $\bm{\theta}$.


The generic estimation problem in \eqref{eq:TOA} will take the form of finding the approximate solution $\bm{\theta}$ to the set of nonlinear simultaneous equations in \eqref{eq:collective1}. However, solving this set of non-linear equations is a difficult task.

We notice that in practice $\hat{\bm{A}}$, $\hat{\bm{y}}$ and $\bm{G}$ are known. If we assume that the measurement noise is small, we can remove $\bm{\eta}_d$ and $\bm{z}$, and approximate \eqref{eq:collective1} by
\begin{align}\label{eq:collective2}
	\hat{\bm{A}}\tilde{\boldsymbol{\theta}}=\hat{\bm{y}}+\bm{G}\left[\lambda_1(\tilde{\boldsymbol{\theta}}),\lambda_2(\tilde{\boldsymbol{\theta}})\right]^T,
\end{align}
where $\tilde{\boldsymbol{\theta}}$ indicates that we have neglected the noise terms $\bm{\eta}_d$ and $\bm{z}$.
If we can find the solution of $\lambda_1$ and $\lambda_2$ from \eqref{eq:collective2}, the position, velocity, clock offset, and clock skew of the moving UN can be estimated.

\subsection{Step 2: Solution for Intermediate Variables}\label{solutionIV}
One approach to solve \eqref{eq:collective2} is to pick an initial guess for $\tilde{\bm{\theta}}$, denoted by $\tilde{\bm{\theta}}^{(0)}$ and estimate $\tilde{\bm{\theta}}$ iteratively as
$$
	\hat{\bm{A}}\tilde{\boldsymbol{\theta}}^{(i)}=\hat{\bm{y}}+\bm{G}\left[\lambda_1(\tilde{\boldsymbol{\theta}}^{(i-1)}),\lambda_2(\tilde{\boldsymbol{\theta}}^{(i-1)})\right]^T.
$$
However, this kind of iterative method is also heavily dependent on the initial guess. In addition, it ignores the structure of $\lambda_1$ and $\lambda_2$ as given by \eqref{eq:lambda}. We aim to propose a direct method to avoid iteration by exploiting the special structure.

Based on (\ref{eq:collective2}), we can roughly estimate $\boldsymbol{\theta}$ with the intermediate variables $\lambda_1$ and $\lambda_2$ by multiplying the pseudo-inverse of $\hat{\bm{A}}$ as
\begin{align}\label{eq:LSestimate}
\tilde{\boldsymbol{\theta}}=\left(\hat{\bm{A}}^T\hat{\bm{A}}\right)^{-1}\hat{\bm{A}}^T\left(\hat{\bm{y}}+\bm{G}\left[\lambda_1(\tilde{\bm{\theta}}),\lambda_2(\tilde{\bm{\theta}})\right]^T\right),
\end{align}
or simply
\begin{align}
\tilde{\boldsymbol{\theta}}=\boldsymbol{g}+\bm{U}\left[\lambda_1(\tilde{\bm{\theta}}),\lambda_2(\tilde{\bm{\theta}})\right]^T,
\end{align}
where $\tilde{\boldsymbol{\theta}}=[\tilde{\boldsymbol{p}}^T,\tilde{\boldsymbol{v}}^T,\tilde{\beta},\tilde{\omega}]^T$,
\begin{align}
\boldsymbol{g}&=\left(\hat{\bm{A}}^T\hat{\bm{A}}\right)^{-1}\hat{\bm{A}}^T\hat{\bm{y}},\\
\bm{U}&=\left(\hat{\bm{A}}^T\hat{\bm{A}}\right)^{-1}\hat{\bm{A}}^T\bm{G}.
\end{align}

To find the solution of the intermediate variables $\lambda_1$ and $\lambda_2$, we design the following two matrices as
\begin{align}
\bm{H}_1&=\left[
\begin{matrix}
\bm{O}_{K\times K} &\bm{O}_{K\times K}&\bm{O}_{K\times 2}\\
\bm{O}_{K\times K} & -\bm{I}_K & \bm{O}_{K\times 2}\\
\bm{O}_{2\times K}  & \bm{O}_{2\times K} &
\begin{matrix}
0&0\\
0&1
\end{matrix}
\end{matrix}
\right],\\
\bm{H}_2&=\left[
\begin{matrix}
	\bm{O}_{K\times K} &-\bm{I}_{K}&\bm{O}_{K\times 2}\\
	-\bm{I}_{K} & \bm{O}_{K\times K} & \bm{O}_{K\times 2}\\
	\bm{O}_{2\times K}  & \bm{O}_{2\times K} &
	\begin{matrix}
		0&1\\
		1&0
	\end{matrix}
\end{matrix}
\right].
\end{align}

To construct equations with respect to the intermediate variables, we use these two matrices $\bm{H}_1$ and $\bm{H}_2$ along with \eqref{eq:lambda} to convert the raw estimate parameter $\tilde{\boldsymbol{\theta}}$ to the linear expression of $\lambda_1$ and $\lambda_2$ as 
\begin{align}
&\tilde{\boldsymbol{\theta}}^T\bm{H}_1\tilde{\boldsymbol{\theta}}=\nonumber\\
&\left(\boldsymbol{g}+\bm{U}\left[\lambda_1(\tilde{\bm{\theta}}),\lambda_2(\tilde{\bm{\theta}})\right]^T\right)^T\bm{H}_1\left(\boldsymbol{g}+\bm{U}\left[\lambda_1(\tilde{\bm{\theta}}),\lambda_2(\tilde{\bm{\theta}})\right]^T\right)\nonumber\\
&=\tilde{\omega}^2-\Vert\tilde{\boldsymbol{v}}\Vert^2=\lambda_1(\tilde{\bm{\theta}}),\label{eq:H1a}\\
&\tilde{\boldsymbol{\theta}}^T\bm{H}_2\tilde{\boldsymbol{\theta}}=\nonumber\\
&\left(\boldsymbol{g}+\bm{U}\left[\lambda_1(\tilde{\bm{\theta}}),\lambda_2(\tilde{\bm{\theta}})\right]^T\right)^T\bm{H}_2\left(\boldsymbol{g}+\bm{U}\left[\lambda_1(\tilde{\bm{\theta}}),\lambda_2(\tilde{\bm{\theta}})\right]^T\right)\nonumber\\
&=2\left(\tilde{\beta}\tilde{\omega}-\tilde{\boldsymbol{p}}^T\tilde{\boldsymbol{v}}\right)= 2\lambda_2(\tilde{\bm{\theta}}). \label{eq:H1b}
\end{align}

After re-organizing (\ref{eq:H1a}) and (\ref{eq:H1b}), we obtain two simultaneous quadratic equations with respect to $\lambda_1$ and $\lambda_2$ as
\begin{align} 
a_1 \lambda_1^2+b_1 \lambda_1\lambda_2 +c_1 \lambda_2^2 + d_1 \lambda_1+e_1 \lambda_2+f_1&=0 \text{,} \label{eq:quadratic1}\\
a_2 \lambda_1^2+b_2 \lambda_1\lambda_2 +c_2 \lambda_2^2 + d_2 \lambda_1+e_2 \lambda_2+f_2&=0 \text{,}\label{eq:quadratic2}
\end{align}
where
$$
a_1=[\bm{U}]_{:,1}^T\bm{H}_1[\bm{U}]_{:,1}, \; b_1=2[\bm{U}]_{:,1}^T\bm{H}_1[\bm{U}]_{:,2},
$$
$$
c_1=[\bm{U}]_{:,2}^T\bm{H}_1[\bm{U}]_{:,2}, \;
d_1=2[\bm{U}]_{:,1}\bm{H}_1\boldsymbol{g}-1,
$$
$$
e_1=2[\bm{U}]_{:,2}\bm{H}_1 \boldsymbol{g},\;
f_1=\boldsymbol{g}^T\bm{H}_1\boldsymbol{g},
$$
$$
a_2=[\bm{U}]_{:,1}^T\bm{H}_2[\bm{U}]_{:,1},\;
b_2=2[\bm{U}]_{:,1}^T\bm{H}_2[\bm{U}]_{:,2},
$$
$$
c_2=[\bm{U}]_{:,2}^T\bm{H}_2[\bm{U}]_{:,2},
d_2=2[\bm{U}]_{:,1}\bm{H}_2\boldsymbol{g},\;
$$
$$
e_2=2[\bm{U}]_{:,2}\bm{H}_2\boldsymbol{g}-2,
f_2=\boldsymbol{g}^T\bm{H}_2\boldsymbol{g},
$$
and $[\cdot]_{:,j}$ represents the $j$-th column of a matrix.

The simultaneous quadratic equations of (\ref{eq:quadratic1}) and (\ref{eq:quadratic2}) can be solved analytically. The detailed approach is given in Appendix \ref{Appendix1}. Note that $\lambda_1$ and $\lambda_2$ are correlated as defined in \eqref{eq:lambda}. However, we ignore this correlation and solve them independently at this step, and rely on the next step to refine the solution. We will show both theoretically and numerically that the refinement step can obtain a final solution that approaches the CRLB under the small noise condition. 

With the solutions of $\lambda_1$ and $\lambda_2$, we can estimate $\bm{\theta}$ based on \eqref{eq:LSestimate}. We notice that $\hat{\bm{A}}^T\hat{\bm{A}}$ must be invertible to ensure the solution of \eqref{eq:LSestimate}. There are $M-1$ rows and $2K+2$ columns in $\hat{\bm{A}}$ as given by \eqref{eq:Amat}. Thus, we need at least $2K+3$ visible ANs or TOA measurements and a proper AN geometry to ensure the full column-rank of $\hat{\bm{A}}$, and the solution of $\tilde{\bm{\theta}}$. Compared with the $2K+5$ visible ANs required by \cite{shi2020sequential}, the new method needs fewer visible ANs.

It is possible that there are multiple roots for $\lambda_1$ and $\lambda_2$, and thus there might be multiple estimates of $\tilde{\boldsymbol{\theta}}$. We use the weighted sum of the squared difference between the observed TOA measurements and the TOA values computed from the solved intermediate variables as a selection criterion for the solution. The selection strategy is given by
\begin{equation} \label{eq:residual}
	\begin{split}
		\min_{\tilde{\boldsymbol{\theta}}} \boldsymbol{r}^T\bm{C}\boldsymbol{r} \text{,}
	\end{split}
\end{equation}
where $\bm{C}$ is the inverse of the covariance matrix for the measurement noise, and
$$
\bm{C}=\mathrm{diag}\left(\frac{1}{\sigma_1^2},\cdots,\frac{1}{\sigma_M^2}\right),
$$
\begin{align} \label{eq:residual1}
[\boldsymbol{r}]_{i}=\hat{\alpha}_i-\left\Vert \tilde{\boldsymbol{p}}+\tilde{\boldsymbol{v}}t_i-\hat{\boldsymbol{p}}_i\right\Vert -\tilde{\beta}-\tilde{\omega}t_i, i=1,\cdots,M.
\end{align}

When the sequential TOA measurement noise variance $\sigma_i$ is identical, the selection strategy (\ref{eq:residual}) can be simplified to the form of $\min_{\tilde{\boldsymbol{\theta}}} \boldsymbol{r}^T\boldsymbol{r}$ to save computation.

\subsection{Step 3: Bias Compensation}
The raw parameter estimate $\tilde{\boldsymbol{\theta}}$ obtained from the previous step needs further refinement to become an optimal estimation. We apply a WLS step using the residuals of all the TOA measurements to refine this raw estimation.

We first express (\ref{eq:TOA}) in the collective form as
\begin{align}\label{eq:allTOA}
\hat{\boldsymbol{\tau}}=h(\boldsymbol{\theta},\bm{p}_\text{AN})+\boldsymbol{\varepsilon},
\end{align}
where $\bm{p}_\text{AN}=[\bm{p}_1^T,\cdots,\bm{p}_M^T]^T$ is the collective form of the true AN positions, $\hat{\boldsymbol{\tau}}=[\hat{\tau}_1,\cdots,\hat{\tau}_M]^T$ is the collective form of the measurements, $h(\boldsymbol{\theta},\bm{p}_\text{AN})$ is a function of $\boldsymbol{\theta}$ and $\bm{p}_\text{AN}$ as given by
$$
\left[h(\boldsymbol{\theta},\bm{p}_\text{AN})\right]_i=\left\Vert\boldsymbol{p}+\boldsymbol{v}t_i-\boldsymbol{p}_i\right\Vert+\beta+\omega t_i-\beta_i,\; i=1,\cdots,M,
$$
and $\boldsymbol{\varepsilon}=[\varepsilon_1,\cdots,\varepsilon_M]^T$.

Note that the raw estimate $\tilde{\boldsymbol{\theta}}$ is not far from the true parameter $\boldsymbol{\theta}$ when the measurement noise and the AN position error are small. Thus, we apply the Taylor series expansion and retain the first order term, and thereby obtain a linear equation as
\begin{align} \label{eq:taylor}
\hat{\boldsymbol{\tau}} &= \mathit{h}(\tilde{\boldsymbol{\theta}},\hat{\bm{p}}_\text{AN}) + \left(\frac{\partial \mathit{h}(\boldsymbol{\theta},\bm{p}_\text{AN})}{\partial \boldsymbol{\theta}}|_{\boldsymbol{\theta}=\tilde{\boldsymbol{\theta}},\bm{p}_\text{AN}=\hat{\bm{p}}_\text{AN}}\right)\left(\boldsymbol{\theta}-\tilde{\boldsymbol{\theta}}\right)\nonumber\\
&+\left(\frac{\partial \mathit{h}(\boldsymbol{\theta},\bm{p}_\text{AN})}{\partial \bm{p}_\text{AN}}|_{\boldsymbol{\theta}=\tilde{\boldsymbol{\theta}},\bm{p}_\text{AN}=\hat{\bm{p}}_\text{AN}}\right)\delta{\boldsymbol{p}}_\text{AN}+\boldsymbol\varepsilon \text{.}
\end{align}
where $\hat{\bm{p}}_\text{AN}=[\hat{\bm{p}}_1^T,\cdots,\hat{\bm{p}}_M^T]^T$ and $\delta{\bm{p}}_\text{AN}=[\delta{\bm{p}}_1^T,\cdots,\delta{\bm{p}}_M^T]^T$ are the collective form of all the known AN positions and their errors, respectively.

We define the design matrix
$$
\tilde{\bm{J}}\triangleq\frac{\partial \mathit{h}(\boldsymbol{\theta},{\bm{p}}_\text{AN})}{\partial \boldsymbol{\theta}}|_{\boldsymbol{\theta}=\tilde{\boldsymbol{\theta}},{\bm{p}}_\text{AN}=\hat{\bm{p}}_\text{AN}},
$$
of which the $i$-th row is given by
\begin{align} \label{eq:Jderivative}
	[\tilde{\bm{J}}]_{i,:}=\left[\frac{\partial \mathit{h}(\boldsymbol{\theta},{\bm{p}}_\text{AN})}{\partial \boldsymbol{\theta}}|_{\boldsymbol{\theta}=\tilde{\boldsymbol{\theta}},{\bm{p}}_\text{AN}=\hat{\bm{p}}_\text{AN}}\right]_{i,:} =\left[-\tilde{\boldsymbol{l}}_{i}^T,-\tilde{\boldsymbol{l}}_{i}^T t_i,1,t_i \right] \text{,}
\end{align}
\begin{equation} \label{eq:LOS}
	\tilde{\boldsymbol{l}}_i=\frac{\hat{\boldsymbol{p}}_i - \tilde{\boldsymbol{p}}-\tilde{\boldsymbol{v}}  t_i}{\Vert \hat{\boldsymbol{p}}_i - \tilde{\boldsymbol{p}}-\tilde{\boldsymbol{v}}  t_i\Vert }, i=1,\cdots,M\text{,}
\end{equation}
and define another matrix
\begin{align}
\tilde{\bm{S}}&\triangleq\frac{\partial \mathit{h}(\boldsymbol{\theta},{\bm{p}}_\text{AN})}{\partial \bm{p}_\text{AN}}|_{\boldsymbol{\theta}=\tilde{\boldsymbol{\theta}},{\bm{p}}_\text{AN}=\hat{\bm{p}}_\text{AN}}=\left[
\begin{matrix}
	\tilde{\bm{l}}_1^T & 0 & \cdots &0\\
	0 &\tilde{\bm{l}}_2^T &  \cdots &0 \\
	\vdots&\vdots &\ddots &\vdots \\
	0 & 0 & \cdots &  \tilde{\bm{l}}_M^T 
\end{matrix}
\right].
\end{align}

We apply the WLS method to (\ref{eq:taylor}) and obtain the refinement vector, denoted by $\Delta \tilde{\boldsymbol{\theta}}$, as
\begin{align} \label{eq:refine1}
\Delta \tilde{\boldsymbol{\theta}}=\left(\tilde{\bm{J}}^T\bm{W}\tilde{\bm{J}}\right)^{-1}\tilde{\bm{J}}^T\bm{W}\left(\hat{\boldsymbol{\tau}}-h(\tilde{\boldsymbol{\theta}},\hat{\bm{p}}_\text{AN})\right),
\end{align}
where the weighting matrix
\begin{align}\label{eq:WC}
\bm{W}&=\left(\mathbb{E}\left[\left(\boldsymbol{\varepsilon}+\tilde{\bm{S}}\delta\bm{p}_{AN}\right)\left(\boldsymbol{\varepsilon}+\tilde{\bm{S}}\delta\bm{p}_{AN}\right)^T\right]\right)^{-1}\nonumber\\
&=\left(\bm{C}^{-1}+\tilde{\bm{S}}\bm{\Sigma}\tilde{\bm{S}}^T\right)^{-1},
\end{align}
and
$$
\bm{\Sigma}=\left[
\begin{matrix}
	\bm{\Sigma}_{1} & &\\
	& \ddots & \\
	&  & \bm{\Sigma}_{M}
\end{matrix}
\right].
$$

Then, the final parameter estimation, denoted by $\tilde{\boldsymbol{\theta}}_{est}=[\tilde{\boldsymbol{p}}_{est}^T,\tilde{\boldsymbol{v}}_{est}^T,\tilde{\beta}_{est},\tilde{\omega}_{est}]^T$, is
\begin{align}\label{eq:refine2}
\tilde{\boldsymbol{\theta}}_{est}=\tilde{\boldsymbol{\theta}}+\Delta\tilde{\boldsymbol{\theta}}.
\end{align}

The procedure of the new CFJLAS method is summarized in Algorithm 1.

\begin{algorithm} 
	\caption{Closed-form JLAS (CFJLAS)}
	\begin{algorithmic}[1]
		\State Input the noisy sequential measurements $\hat{{\tau}}_i$, and the known AN positions $\hat{\boldsymbol{p}}_{i}$ and clock offsets $\beta_{i}$,  $ i=1,\cdots,M$.
		\State Step 1: Reparameterization: Square and difference the TOA measurement equations, and form matrices $\hat{\bm{A}}$, $\bm{G}$ and vector $\hat{\bm{y}}$ based on (\ref{eq:Amat}), \eqref{eq:matG} and \eqref{eq:vectory}.
		\State Step 2: Solution of intermediate variables: Solve the quadratic equations (\ref{eq:quadratic1}) and (\ref{eq:quadratic2}) and select the root based on (\ref{eq:residual}) to obtain the raw parameter estimate $\tilde{\boldsymbol{\theta}}$.
		\State Step 3: Bias compensation: Compute the refined parameter $\tilde{\boldsymbol{\theta}}_{est}$ using (\ref{eq:refine1}) and (\ref{eq:refine2}).
		\State Output the parameter estimate result $\tilde{\boldsymbol{\theta}}_{est}$.
	\end{algorithmic}
\end{algorithm}

We can see from the steps of the new CFJLAS method that we exploit the structure of the particular pair of nonlinear equations. By doing so, we have developed a closed-form solution, which solves the two intermediate variables analytically and then computes the parameters using the intermediate variables. Compared with the method in \cite{shi2020sequential}, which jointly estimates the newly employed variables and the parameters, the new CFJLAS requires fewer measurements by nature since it does not need to estimate the intermediate variables along with the parameters.

\section{Performance Analysis} \label{locanalysis}
It is of significant interest to evaluate the achievable estimation performance of this new CFJLAS approach. In this section, we first derive the CRLB, which is a lower bound on the achievable estimation error variance to quantify the estimation performance. After that, a theoretical error analysis is conducted to evaluate the estimation accuracy of the proposed CFJLAS compared with the CRLB. Then, we analyze the computational complexity in comparison with the iterative ML method (LSPM-UVD) presented by \cite{zhao2020optloc}.

\subsection{CRLB Derivation}\label{CRLBanalysis}
CRLB is the lower bound of the estimation accuracy of an unbiased estimator. We use the CRLB as the benchmark for the optimal JLAS estimation \cite{vaghefi2015cooperative}. We first derive the Fisher information matrix (FIM), denoted by $\bm{\mathcal{F}}$, to obtain the CRLB.

We consider the general case that the known AN positions have errors. Let $\bm{\zeta}=[\bm{\theta}^T,\bm{p}_\text{AN}^T]^T$. Then the likelihood function, denoted by $p(\hat{\boldsymbol{\tau}},\hat{\bm{p}}_\text{AN}|\bm{\zeta})$, is
\begin{align} \label{eq:likelihood}
	&p(\hat{\boldsymbol{\tau}},\hat{\bm{p}}_\text{AN}|\bm{\zeta}) = \frac{\exp\left(-\frac{1}{2} \left(\hat{\boldsymbol{\tau}} - \mathit{h}({\boldsymbol{\theta}})\right)^T\bm{C}\left(\hat{\boldsymbol{\tau}} - \mathit{h}({\boldsymbol{\theta}})\right)\right)}{(2\pi)^{\frac{M}{2}} |\bm{C}|^{-\frac{1}{2}}}\nonumber\\
	&\cdot \frac{\exp\left(-\frac{1}{2} \left(\hat{\boldsymbol{p}}_\text{AN} - \boldsymbol{p}_\text{AN}\right)^T\bm{\Sigma}^{-1}\left(\hat{\boldsymbol{p}}_\text{AN} - \boldsymbol{p}_\text{AN}\right)\right)}{(2\pi)^{\frac{KM}{2}} |\bm{\Sigma}|^{\frac{1}{2}}}
	 \text{.}
\end{align}

The FIM for $\bm{\zeta}$ is then given by
\begin{align} \label{eq:FisherExpectation0}
	\bm{\mathcal{F}}(\bm{\zeta})&=-\mathbb{E}\left[\frac{\partial^2 \ln p(\hat{\boldsymbol{\tau}},\hat{\bm{p}}_\text{AN}|\bm{\zeta})}{\partial\boldsymbol{\zeta} \partial\boldsymbol{\zeta}^T}\right] \nonumber\\
	&=
	\left[\begin{matrix}
		\bm{J}^T\bm{C}\bm{J} & \bm{J}^T\bm{C}\bm{S}\\
		\bm{S}^T\bm{C}\bm{J}&\bm{S}^T\bm{C}\bm{S}+\bm{\Sigma}^{-1}
	\end{matrix}
	\right]
	\text{,}
\end{align}
where
\begin{align}
[{\bm{J}}]_{i,:}=\left[\frac{\partial \mathit{h}(\boldsymbol{\theta})}{\partial \boldsymbol{\theta}}\right]_{i,:} =\left[-{\boldsymbol{l}}_{i}^T,-{\boldsymbol{l}}_{i}^T t_i,1,t_i \right],
\end{align}
\begin{align}
{\boldsymbol{l}}_i=\frac{\boldsymbol{p}_i - {\boldsymbol{p}}-{\boldsymbol{v}}  t_i}{\Vert \boldsymbol{p}_i - {\boldsymbol{p}}-{\boldsymbol{v}}  t_i\Vert }, i=1,\cdots,M\text{.}
\end{align}
\begin{align}
	\bm{S}=\left[
	\begin{matrix}
		\bm{l}_1^T & 0 & \cdots &0\\
		0 &\bm{l}_2^T &  \cdots &0 \\
		\vdots&\vdots &\ddots &\vdots \\
		0 & 0 & \cdots &  \bm{l}_M^T 
	\end{matrix}
	\right].
\end{align}

The CRLB for $\bm{\theta}$ is the top-left $(2K+2)\times(2K+2)$ submatrix of $\bm{\mathcal{F}}^{-1}(\bm{\zeta})$, i.e.,
\begin{align} \label{eq:CRLB_Fisher}
	&\mathsf{CRLB}(\bm{\theta})=\left[\bm{\mathcal{F}}^{-1}(\boldsymbol{\zeta})\right]_{1:2K+2,1:2K+2}=\nonumber\\
	& \left(\bm{J}^T\bm{C}\bm{J}-\bm{J}^T\bm{C}\bm{S}(\bm{S}^T\bm{C}\bm{S}+\bm{\Sigma}^{-1})^{-1}\bm{S}^T\bm{C}\bm{J}\right)^{-1}.
\end{align}

\subsection{Theoretical Error Analysis} \label{ErrorAnalysis}

In this subsection, we investigate the theoretical error of the new CFJLAS.

We first look into the estimation error of Step 2: Solution for Intermediate Variables. We consider the ideal case with no AN position error or TOA measurement noise and have the following relation with a similar expression to \eqref{eq:collective2} as
\begin{align}\label{eq:noisefree}
	{\bm{A}}{\boldsymbol{\theta}}={\bm{y}}+\bm{G}\left[\lambda_1({\boldsymbol{\theta}}),\lambda_2({\boldsymbol{\theta}})\right]^T,
\end{align}
where $\bm{A}$ and $\bm{y}$ are the error-free version of $\hat{\bm{A}}$ and $\hat{\bm{y}}$, respectively.

We denote the estimation error by $\Delta \bm{\theta}=[\Delta\bm{p}^T,\Delta\bm{v}^T,\Delta\omega,\Delta\beta]^T$, the errors due to the AN position error and measurement noise in $\bm{A}$, $\bm{y}$, $\lambda_1$ and $\lambda_2$ by $\Delta\bm{A}$, $\Delta\bm{y}$, $\Delta\lambda_1$ and $\Delta\lambda_2$, respectively. Under the small noise condition, we ignore the second order terms of the AN position error and measurement noise, and have
\begin{align}\label{eq:noisyrelation}
	&\left({\bm{A}}+\Delta\bm{A}\right)\left({\bm{\theta}}+\Delta \bm{\theta}\right)=\nonumber\\
	&\left({\bm{y}}+\Delta\bm{y}\right)+\bm{G}\left[\lambda_1({\bm{\theta}})+\Delta\lambda_1({\bm{\theta}}),\lambda_2({\bm{\theta}})+\Delta\lambda_2({\bm{\theta}})\right]^T,
\end{align}
where
\begin{align}
	&\left[\Delta\bm{A}\right]_{i,:}=\nonumber\\
	&2\left[
		\delta\bm{p}_i^T-\delta\bm{p}_1^T, t_i\delta\bm{p}_i^T-t_1\delta\bm{p}_1^T, \varepsilon_1-\varepsilon_i, t_1\varepsilon_1-t_i\varepsilon_i,
	\right],\nonumber\\
&\left[\Delta\bm{y}\right]_{i}=
2\left(\bm{p}_i^T\delta\bm{p}_i-\bm{p}_1^T\delta\bm{p}_1\right)-2\left({\alpha}_i\varepsilon_i-{\alpha}_1\varepsilon_i\right),\nonumber\\
&i=2,\cdots,M,\nonumber\\
&\Delta \lambda_1(\bm{\theta})=2\bm{\theta}^T\bm{H}_1
\Delta\bm{\theta}+\Delta\bm{\theta}^T\bm{H}_1
\Delta\bm{\theta},\nonumber\\
&\Delta \lambda_2(\bm{\theta})=\bm{\theta}^T\bm{H}_2
\Delta\bm{\theta}+\Delta\bm{\theta}^T\bm{H}_2
\Delta\bm{\theta}/2,
\end{align}
and $\alpha_i$ is the error-free version of $\hat{\alpha}_i$.

We subtract \eqref{eq:noisefree} from \eqref{eq:noisyrelation}, and come to
\begin{align}\label{eq:errorstep1}
	&\left(\bm{A}+\Delta \bm{A}-\bm{G}\left[2\bm{H}_1 \boldsymbol{\theta},\bm{H}_2 \boldsymbol{\theta}\right]^T\right)\Delta \boldsymbol{\theta} \nonumber\\
	&=\bm{G}\left[\Delta \boldsymbol{\theta}^T\bm{H}_1\Delta\boldsymbol{\theta},\Delta \boldsymbol{\theta}^T\bm{H}_2\Delta\boldsymbol{\theta}/2\right]^T+\Delta \boldsymbol{y}-\Delta \bm{A}\boldsymbol{\theta}.
\end{align}

We show from \eqref{eq:errorstep1} that the estimation error $\Delta\bm{\theta}$ is related to $\Delta \bm{A}$, $\Delta \bm{y}$ and $\bm{\theta}$, and is difficult to be obtained analytically. We take the expectation for both sides of \eqref{eq:errorstep1} and have
\begin{align}\label{eq:Eerrorstep1}
	&\left(\bm{A}-\bm{G}\left[2\bm{H}_1 \boldsymbol{\theta},\bm{H}_2 \boldsymbol{\theta}\right]^T\right)\mathbb{E}[\Delta \boldsymbol{\theta}] \nonumber\\
	&=\bm{G}\mathbb{E}\left[\Delta \omega^2-\Vert\Delta\bm{v}\Vert^2,\Delta {\beta}\Delta\omega-\Delta \bm{p}^T\Delta\bm{v}\right]^T.
\end{align}

We show from \eqref{eq:Eerrorstep1} that the estimation bias $\mathbb{E}[\Delta \boldsymbol{\theta}]$ is generally not zero, indicating a biased estimation in this step. This is still a complex relation and is difficult to obtain an analytical solution. Furthermore, numerical simulations in Section \ref{simulation} shows that the estimation error of Step 2 deviates from the CRLB or is biased. However, the error will reduce significantly and become sufficiently small when the number of ANs increases, as shown in Table \ref{table_stats}, and thus ensure the asymptotically optimal solution of Step 3.

We now study the estimation error of Step 3: Bias Compensation. Introducing (\ref{eq:allTOA}), we re-write (\ref{eq:refine1}) as
\begin{align} \label{eq:expand1}
\Delta \tilde{\boldsymbol{\theta}}=&\left(\tilde{\bm{J}}^T\bm{W}\tilde{\bm{J}}\right)^{-1}\tilde{\bm{J}}^T\bm{W}\left(\hat{\boldsymbol{\tau}}-h(\tilde{\boldsymbol{\theta}},\hat{\bm{p}}_\text{AN})\right) \nonumber\\
=&\left(\tilde{\bm{J}}^T\bm{W}\tilde{\bm{J}}\right)^{-1}\tilde{\bm{J}}^T\bm{W}\cdot\nonumber\\
&\left(\hat{\boldsymbol{\tau}}-h({\boldsymbol{\theta}},{\bm{p}}_\text{AN})+h({\boldsymbol{\theta}},{\bm{p}}_\text{AN})-h(\tilde{\boldsymbol{\theta}},\hat{\bm{p}}_\text{AN})\right)\nonumber\\
=&\left(\tilde{\bm{J}}^T\bm{W}\tilde{\bm{J}}\right)^{-1}\tilde{\bm{J}}^T\bm{W}\boldsymbol{\varepsilon}+\nonumber\\
&\left(\tilde{\bm{J}}^T\bm{W}\tilde{\bm{J}}\right)^{-1}\tilde{\bm{J}}^T\bm{W}\left(h({\boldsymbol{\theta}},{\bm{p}}_\text{AN})-h(\tilde{\boldsymbol{\theta}},\hat{\bm{p}}_\text{AN})\right).
\end{align}

According to (\ref{eq:allTOA}) and (\ref{eq:taylor}), we have
\begin{align}\label{eq:ftheta}
\mathit{h}(\boldsymbol{\theta},{\bm{p}}_\text{AN})= \mathit{h}(\tilde{\boldsymbol{\theta}},\hat{\bm{p}}_\text{AN}) + \tilde{\bm{J}}\cdot\left(\boldsymbol{\theta}-\tilde{\boldsymbol{\theta}}\right)+\tilde{\bm{S}}\cdot\delta{\boldsymbol{p}}_\text{AN}.
\end{align}

By substituting (\ref{eq:ftheta}) into (\ref{eq:expand1}), we come to
\begin{align} \label{eq:dtheta}
\Delta \tilde{\boldsymbol{\theta}}=
&\left(\tilde{\bm{J}}^T\bm{W}\tilde{\bm{J}}\right)^{-1}\tilde{\bm{J}}^T\bm{W}\boldsymbol{\varepsilon}+\nonumber\\
&\left(\tilde{\bm{J}}^T\bm{W}\tilde{\bm{J}}\right)^{-1}\tilde{\bm{J}}^T\bm{W}\tilde{\bm{J}}\left({\boldsymbol{\theta}}-\tilde{\boldsymbol{\theta}}\right)+\nonumber\\
&\left(\tilde{\bm{J}}^T\bm{W}\tilde{\bm{J}}\right)^{-1}\tilde{\bm{J}}^T\bm{W}\tilde{\bm{S}}\cdot\delta\bm{p}_\text{AN}\nonumber\\
=&\left(\tilde{\bm{J}}^T\bm{W}\tilde{\bm{J}}\right)^{-1}\tilde{\bm{J}}^T\bm{W}\left(\boldsymbol{\varepsilon}+\tilde{\bm{S}}\cdot\delta\bm{p}_\text{AN}\right)+{\boldsymbol{\theta}}-\tilde{\boldsymbol{\theta}}.
\end{align}

Based on (\ref{eq:refine2}), we re-organize (\ref{eq:dtheta}) as 
\begin{align}
\tilde{\boldsymbol{\theta}}_{est}-{\boldsymbol{\theta}}=\left(\tilde{\bm{J}}^T\bm{W}\tilde{\bm{J}}\right)^{-1}\tilde{\bm{J}}^T\bm{W}\left(\boldsymbol{\varepsilon}+\tilde{\bm{S}}\cdot\delta\bm{p}_\text{AN}\right).
\end{align}

The error expectation or estimation bias is 
\begin{align} \label{eq:Estep3}
	\mathbb{E}\left[\tilde{\boldsymbol{\theta}}_{est}-{\boldsymbol{\theta}}\right]=\mathbb{E}\left[\left(\tilde{\bm{J}}^T\bm{W}\tilde{\bm{J}}\right)^{-1}\tilde{\bm{J}}^T\bm{W}\left(\boldsymbol{\varepsilon}+\tilde{\bm{S}}\delta\bm{p}_\text{AN}\right)\right].
\end{align}
	
Taking \eqref{eq:WC} into account, we obtain the covariance of the estimation error as
\begin{align}\label{eq:errcov}
\mathbb{E}&\left[\left(\tilde{\boldsymbol{\theta}}_{est}-{\boldsymbol{\theta}}\right)\left(\tilde{\boldsymbol{\theta}}_{est}-{\boldsymbol{\theta}}\right)^T\right]\nonumber\\
=&\left(\tilde{\bm{J}}^T\bm{W}\tilde{\bm{J}}\right)^{-1}\tilde{\bm{J}}^T\bm{W} \cdot \nonumber\\
&\mathbb{E}\left[\left(\boldsymbol{\varepsilon}+\tilde{\bm{S}}\cdot\delta\bm{p}_\text{AN}\right)\left(\boldsymbol{\varepsilon}+\tilde{\bm{S}}\cdot\delta\bm{p}_\text{AN}\right)^T\right]\bm{W}\tilde{\bm{J}}\left(\tilde{\bm{J}}^T\bm{W}\tilde{\bm{J}}\right)^{-1}\nonumber\\
=&\left(\tilde{\bm{J}}^T\bm{W}\tilde{\bm{J}}\right)^{-1}\tilde{\bm{J}}^T\bm{W} \mathbb{E}\left[\boldsymbol{\varepsilon}\boldsymbol{\varepsilon}^T+\tilde{\bm{S}}\delta\bm{p}_\text{AN}\delta\bm{p}_\text{AN}^T\tilde{\bm{S}}^T\right]\cdot\nonumber\\
&\bm{W}\tilde{\bm{J}}\left(\tilde{\bm{J}}^T\bm{W}\tilde{\bm{J}}\right)^{-1}\nonumber\\
=&\left(\tilde{\bm{J}}^T\bm{W}\tilde{\bm{J}}\right)^{-1}\tilde{\bm{J}}^T\bm{W}\left(\bm{C}^{-1}+\tilde{\bm{S}}\bm{\Sigma}\tilde{\bm{S}}^T\right)\bm{W}\tilde{\bm{J}}\left(\tilde{\bm{J}}^T\bm{W}\tilde{\bm{J}}\right)^{-1}\nonumber\\
=&\left(\tilde{\bm{J}}^T\bm{W}\tilde{\bm{J}}\right)^{-1}.
\end{align}

We expand the inverse in \eqref{eq:WC} as
\begin{align}\label{eq:Winverse}
	\bm{W}&=\bm{C}-\bm{C}\tilde{\bm{S}}^T(\tilde{\bm{S}}^T\bm{C}\tilde{\bm{S}}+\bm{\Sigma}^{-1})^{-1}\tilde{\bm{S}}^T\bm{C}
	,
\end{align}
plug it into \eqref{eq:errcov}, and come to
\begin{align}\label{eq:errcov2}
	&\mathbb{E}\left[\left(\tilde{\boldsymbol{\theta}}_{est}-{\boldsymbol{\theta}}\right)\left(\tilde{\boldsymbol{\theta}}_{est}-{\boldsymbol{\theta}}\right)^T\right]=\nonumber\\
	&\left(\tilde{\bm{J}}\bm{C}\tilde{\bm{J}}^{T}-\tilde{\bm{J}}\bm{C}\tilde{\bm{S}}(\tilde{\bm{S}}^T\bm{C}\tilde{\bm{S}}+\bm{\Sigma}^{-1})^{-1}\tilde{\bm{S}}^T\bm{C}\tilde{\bm{J}}\right)^{-1}.
\end{align}

Note that when the noise becomes small, $\tilde{\bm{J}}$ is asymptotically the same as the true ${\bm{J}}$. Similarly,  $\tilde{\bm{S}}$ is essentially the same as ${\bm{S}}$ when the AN position error is small. Therefore, the estimation bias $\mathbb{E}\left[\tilde{\boldsymbol{\theta}}_{est}-{\boldsymbol{\theta}}\right]$ given by \eqref{eq:Estep3} goes to zero, and the error covariance given by (\ref{eq:errcov2}) reaches the CRLB given by (\ref{eq:CRLB_Fisher}), under the small noise and small AN error condition.

\subsection{Complexity Analysis} \label{comana}



Following \cite{golub2013matrix}, a floating point addition, multiplication or square root operation for real numbers can be done in one flop. In this subsection, we estimate the complexities of the CFJLAS and the iterative method by counting the flops of the major operations.

We denote the total complexity of the CFJLAS by $D$. As part of $D$, the complexity caused by the step of solving the equations \eqref{eq:quadratic1} and \eqref{eq:quadratic2} is fixed, not related to the position dimension $K$ or the number of ANs $M$. Other complexities lie in finding the proper roots in \eqref{eq:residual}, raw estimate in \eqref{eq:LSestimate} and WLS refinement in \eqref{eq:refine1}. The detailed complexity analysis is given in Appendix \ref{Appendix2}. We count the total number of flops as approximated by
\begin{align} \label{eq:complexD}
D \approx& 32K^3 +16K^2M + 104K^2+ 62KM +\nonumber\\
&148K + 65M + 697,
\end{align}
where $K$ is the dimension of the position, and $M$ is the number of ANs.

For the iterative ML method (LSPM-UVD) in \cite{zhao2020optloc}, the total complexity is determined by the computation in each iteration, denoted by $L$, and the number of iterations, denoted by $n$. The major operations in each iteration are comprised of matrix multiplication and matrix inversion. Each iteration has the same complexity. Thus, the total complexity of the iterative method is $nL$. The detailed analysis is presented in Appendix \ref{Appendix2}. The number of flops for $L$ is approximated by
\begin{align}\label{eq:complexL}
L\approx16K^3 + 8K^2M+56K^2 + 22KM+64K + 16M + 24.
\end{align}

We can see that the proposed new CFJLAS method has a complexity of $D$, which is the function of the problem size $K$ and $M$. For the iterative method, the complexity is $nL$, which will grow as the number of iterations $n$ increases in addition to the problem size $K$ and $M$. In other words, given the problem size, the complexity of the new CFJLAS is fixed while that of the iterative method will grow with an increasing number of iterations. This indicates a more controllable and predictable complexity of the new CFJLAS than the conventional iterative method.

Furthermore, we substitute $K=2$ and $M=8$ into \eqref{eq:complexD} and \eqref{eq:complexL} and obtain that $D=3689$ and $L=1240$. We can see that $L$ and $D/3$ are approximately the same. Therefore, we can expect that the new CFJLAS method has lower computational complexity when the number of iteration exceeds 3 for the iterative method. This will be shown by simulation in Section V.

\section{Numerical Simulation} \label{simulation}
In this section, numerical simulations are carried out to evaluate the performance of the new CFJLAS method. The iterative ML method (LSPM-UVD) \cite{zhao2020optloc} is selected for comparison. The computational platform running the following simulations is Matlab R2019b on a PC with Intel Core i5-8400 CPU @2.8GHz and 32G RAM.

\subsection{Performance Metrics}
The root mean square error (RMSE) is used to evaluate the performance in the following numerical simulations. The RMSEs of the position $\boldsymbol{p}$, velocity $\boldsymbol{v}$, clock offset $\beta$ and clock skew $\omega$ are given by
\begin{align}\label{eq:MSEerr}
\text{RMSE}_{\boldsymbol{p}}&=\sqrt{\frac{1}{N}\sum_{1}^{N}\Vert\boldsymbol{p}-\tilde{\boldsymbol{p}}_{est}\Vert^2}, \nonumber\\
\text{RMSE}_{\boldsymbol{v}}&=\sqrt{\frac{1}{N}\sum_{1}^{N}\Vert\boldsymbol{v}-\tilde{\boldsymbol{v}}_{est}\Vert^2},\nonumber\\
\text{RMSE}_{\beta}&=\sqrt{\frac{1}{N}\sum_{1}^{N}\Vert{\beta}-\tilde{\beta}_{est}\Vert^2},\nonumber\\
\text{RMSE}_{\omega}&=\sqrt{\frac{1}{N}\sum_{1}^{N}\Vert\omega-\tilde{\omega}_{est}\Vert^2},
\end{align}
where $N$ is the total number of Monte-Carlo runs, the variables with ``$\sim$'' overhead are the estimated parameters from the proposed CFJLAS method.

CRLB is used as a benchmark to evaluate the lower bound of the accuracy. In the simulation scene, the theoretical lower bound of the position, the velocity, the clock offset and clock skew are computed by
\begin{align} \label{eq:errCRLB}
\text{error}_{\boldsymbol{p}}=&\sqrt{\frac{1}{N}\sum_{1}^{N}\sum_{i=1}^{K}\left[\mathsf{CRLB}\right]_{i,i}}, \nonumber\\
\text{error}_{\boldsymbol{v}}=&\sqrt{\frac{1}{N}\sum_{1}^{N}\sum_{i=K+1}^{2K}\left[\mathsf{CRLB}\right]_{i,i}}, \nonumber\\
\text{error}_{\beta}=&\sqrt{\frac{1}{N}\sum_{1}^{N}\left[\mathsf{CRLB}\right]_{2K+1,2K+1}}, \nonumber\\
\text{error}_{\omega}=&\sqrt{\frac{1}{N}\sum_{1}^{N}\left[\mathsf{CRLB}\right]_{2K+2,2K+2}},
\end{align}
where $[\cdot]_{i,j}$ is the entry at the $i$-th row and the $j$-th column of a matrix.

\subsection{Simulation Settings} \label{simsetup}

\begin{figure}
	\centering
	\includegraphics[width=0.99\linewidth]{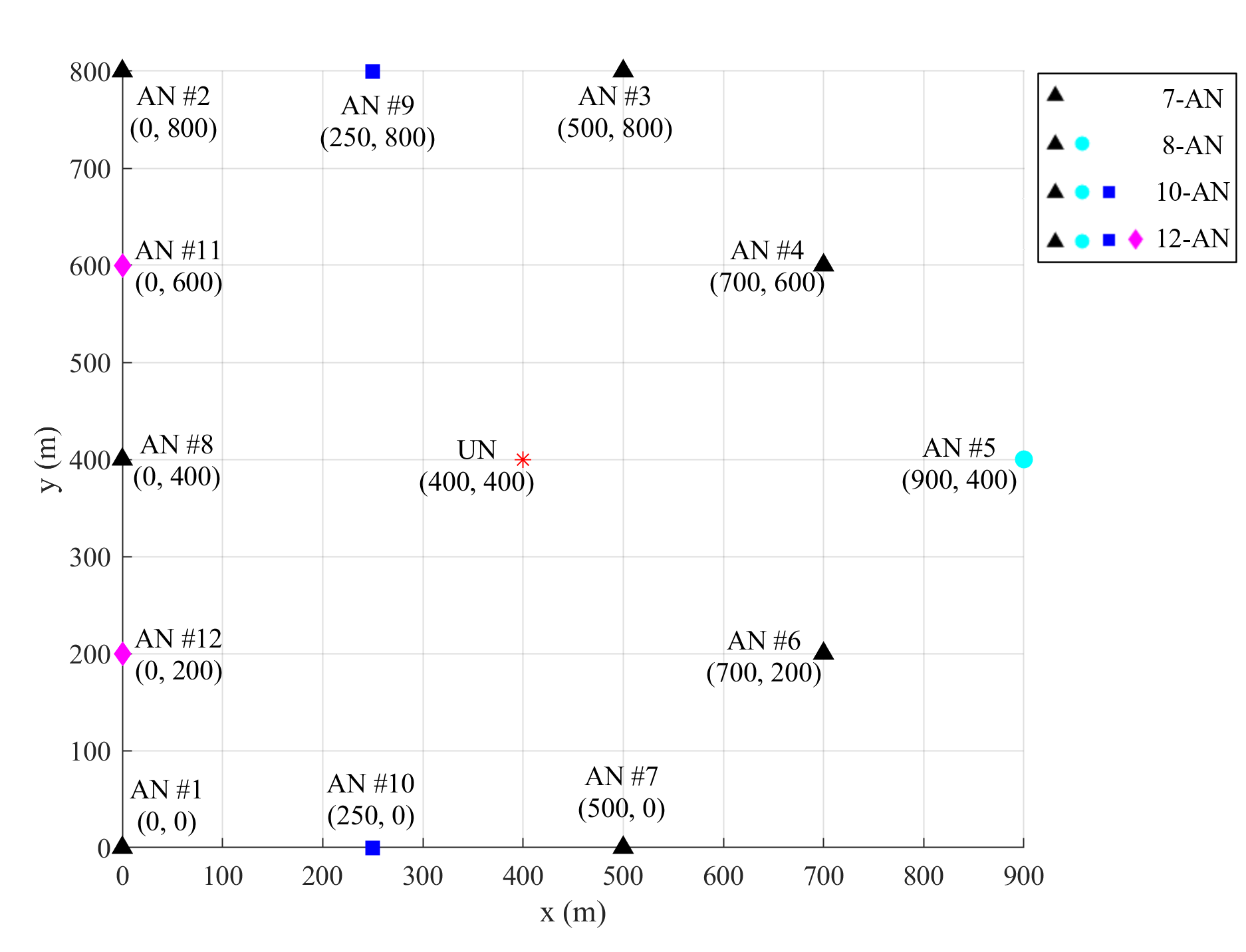}
	\caption{ ANs and UN in the simulation scene. The combinations of the ANs used in the four different simulation situations are shown in the figure.}
	\label{fig:simulationsetting}
\end{figure}

We create a 2D simulation scene with a JLAS system comprised of 12 ANs and 1 UN as shown by Fig. \ref{fig:simulationsetting}. It resembles the drone formation flight in the real world. The ANs represent the drones that form the relative spatiotemporal coordinate system and the UN represent the target to be located. We consider four different situations with 7, 8, 10 and 12 ANs in the following subsection, as given by the figure. The stand deviation (STD) of the $x$ and $y$-axis errors of all the known AN positions are set to 0.5 m. The UN is initially placed near the center of this formation. The relative speed between the UD and ANs are randomly set from $ \mathcal{U}(0,50)$ m/s, which is beyond the motion of a consumer-level quadrotor. We allocate different time slots to the ANs with an interval of 5 ms between successive message transmissions. The initial clock offset and initial clock skew of the UN for each simulation run are set as random variables, drawn from uniform distributions, i.e., $\beta\sim \mathcal{U}(-10^{-5},10^{-5})$ s, and $\omega\sim \mathcal{U}(-20,20)$ part per million (ppm), respectively.

For the iterative method \cite{zhao2020optloc}, the maximum number of iterations is set to 10 and the convergence threshold is set to 0.01 m, i.e., the iteration stops if the increment of the $(k+1)$-th iteration $\Vert[\tilde{\boldsymbol{p}}_{est}^T(k+1),\tilde{\beta}_{est}(k+1)]^T-[\tilde{\boldsymbol{p}}_{est}^T(k),\tilde{\beta}_{est}(k)]^T\Vert<0.01$. In each iteration, a matrix inversion is conducted to estimate the parameter increment. Before inversion, we compute the reciprocal condition number ($rcond$) to determine if the matrix is singular or not. Specifically, we set the threshold of the reciprocal condition number to $10^{-15}$, i.e., if $rcond<10^{-15}$, the matrix is considered singular and the iteration stops.

\subsection{Estimation Performance with Different Numbers of ANs}

We simulate four situations, i.e., 7 ANs (black-triangle ANs in Fig. \ref{fig:simulationsetting}), 8 ANs (7-AN plus the cyan-dot AN), 10 (8-AN plus blue-square ANs) and 12 ANs (10-AN plus magenta-diamond ANs). In this 2D scene, $2K+3=7$ is the minimum number of measurements required by the new CFJLAS method. We set the TOA measurement noise $\sigma$ varying from 0.1 m to 10 m with a step of 1.1 m. At each step, we conduct 100,000 Monte-Carlo simulations.

\begin{figure}
	\centering
	\includegraphics[width=0.99\linewidth]{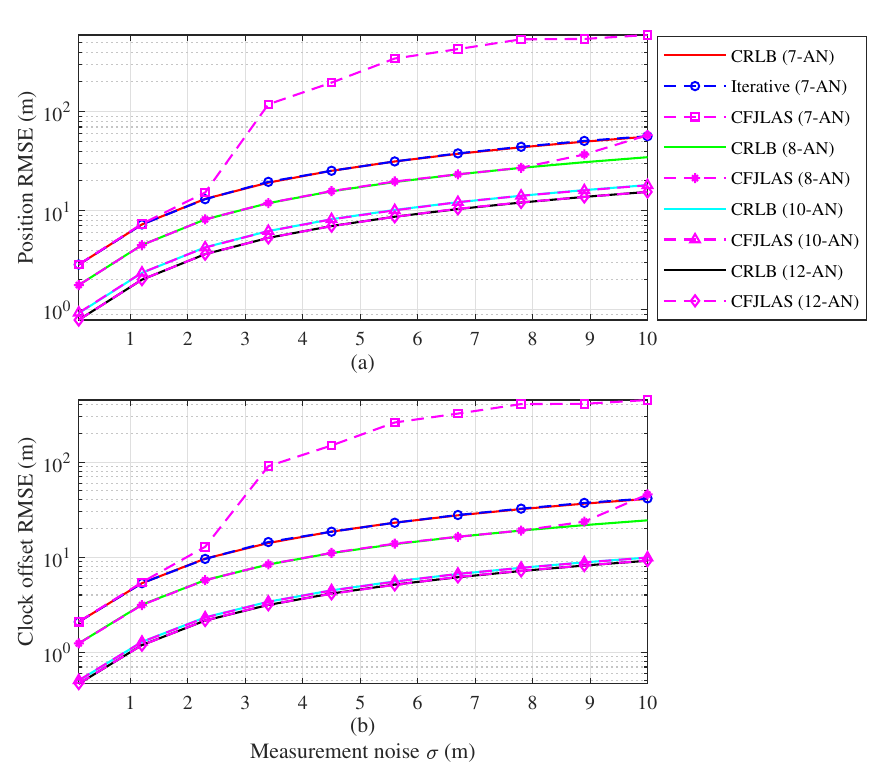}
	\caption{Estimation RMSEs vs. measurement noise. The iterative method is initialized with the ground truth to ensure the optimal estimation result. The RMSEs of the new CFJLAS method reach the CRLB under the small noise condition. With an increasing number of ANs, the new CFJLAS method provides smaller estimation errors even when the measurement noise grows.}
	\label{fig:perrorvsnoise}
\end{figure}

Fig. \ref{fig:perrorvsnoise} illustrates the estimation error results with varying measurement noise. The theoretical CRLB is computed based on (\ref{eq:errCRLB}), and the RMSEs are the average of 100,000 Monte-Carlo runs based on (\ref{eq:MSEerr}). For the iterative method, to obtain the optimal ML result, the initial guess of the UN's parameters is set to the ground truth. The results of the iterative method reach the CRLB in all the situations with different numbers of ANs and thus are not shown for the situations with more than 7 ANs for better illustration.

We take Fig. \ref{fig:perrorvsnoise} (a) as an example. It can be seen that in the 7-AN situation, the positioning accuracy of the proposed CFJLAS method reaches the CRLB when the measurement noise is under 1.2 m. When the measurement noise becomes large, the positioning error will deviate from the CRLB and the new CFJLAS method may produce worse solutions. With an increasing number of ANs to 8, 10 and 12, the position error of the new CFJLAS remains close to the CRLB within a larger range of the measurement noise. In addition, with more ANs, the position error becomes smaller given the same measurement noise. The clock errors have the same pattern as the position errors. We also verified that the estimation errors for the velocity and clock skew have the same pattern, which are not shown for better illustration. These results show that i) with the minimal number of ANs, the new CFJLAS method can achieve the CRLB under a very small noise condition, and ii) increasing the number of ANs can significantly reduce the estimation errors and keep the errors close to the CRLB when the measurement noise increases.



In order to evaluate how the CFJLAS' intermediate steps work and examine the measures of the variance, we take a close look at some error statistics for each situation. We select the 90 percentile, 10 percentile and RMSE for the four situations with 7, 8, 10, and 12 ANs when the measurement noise $\sigma=5.6$ m, and list them in Table \ref{table_stats}. In addition, we use a threshold of $3\sqrt{\mathsf{CRLB}}$ to determine if the localization result is correct or not, i.e., if the estimated position error is smaller than $3\sqrt{\mathsf{CRLB}}$, the estimation result is determined to be correct.

As can be seen from Table \ref{table_stats}, the RMSEs of Step 2 are larger than the CRLB, showing the biased estimation in Step 2. With more ANs, the errors including the 90 percentile, 10 percentile and the RMSE of Step 2 and the final results decrease, and the correctness rate increases, showing an improvement in the estimation performance. From the 90 percentile errors, and the correctness rate, we can see that there is a small probability that the new CFJLAS method produces large error results, even in the situation with the minimal 7 ANs. This demonstrates the effectiveness of the new method in JLAS estimation.

\begin{table}[!t]
	\centering
	\begin{threeparttable}
		\caption{Position Error Statistics for CFJLAS with Different Number of ANs}
		\label{table_stats}
		\centering
		\begin{tabular}{c c r r r r }
			\toprule
			\multicolumn{2}{c}{Number of ANs} & 7 & 8 & 10 & 12 \\
			\hline
			\multirow{3}{0.8cm}{Step 2}
			& 90 percentile (m) & 245.59	&35.93	&19.08	&14.49	\\
			& 10 percentile (m) & 10.84	&5.06	&3.38	&2.84	\\
			& RMSE (m) & 357.93	&22.40	&12.55	&9.58	\\
			\cline{1-6}
			\multirow{3}{0.8cm}{Final}
			& 90 percentile (m) & 59.35	&31.71	&15.74	&13.39	\\
			& 10 percentile (m) & 7.57	&4.83	&2.95	&2.54	\\
			& RMSE (m) & 344.15	&19.71	&10.18	&8.67	\\
			\cline{1-6}
			\multicolumn{2}{c}{$\sqrt{\mathsf{CRLB}}$ (m)} &31.39 &19.46 &10.17 &8.67\\
			\cline{1-6}
			\multicolumn{2}{c}{Correctness rate (\%)} &98.30 &99.76 &99.92 &99.92\\
			\bottomrule
		\end{tabular}
		\begin{tablenotes}[para,flushleft]
			Note: In all four situations with different number of ANs, the measurement noise is set to $\sigma=5.6$ m. The correctness rate is the percentage of the simulation runs in which the final position error$<3\sqrt{\mathsf{CRLB}}$. The 90 percentile, 10 percentile errors and RMSEs of Step 2 are larger than those of the final results and the CRLB, showing a biased estimation in Step 2. With an increasing number of ANs, this RMSE decreases dramatically. The 90 percentile errors of both Step 2 and the final results as well as the correctness rate show that the new CFJLAS has a small probability to generate large error results.
		\end{tablenotes}
	\end{threeparttable}
\end{table}

\subsection{Comparison with Iterative Method}
Note that the results of the iterative method shown in Fig. \ref{fig:perrorvsnoise} is in the ideal case with ground truth initialization. In practice, we do not know the true values of the parameters to be estimated and thus cannot initialize the iterative method with the ground truth. 

To compare the performance of the two methods, we initialize the iterative method by different position errors with STD=10, 50, 100, 150 and 200 m. We use the 8-AN case for this simulation. We run 100,000 Monte-Carlo simulations at each measurement noise step. Other settings remain the same.

We note from the simulation settings that there are three cases for the iterative method to stop, i.e.,
	
\hspace{-3mm}(a) \textit{Convergence}: the norm of the parameter estimation increment in the current iteration is smaller than 0.01 m,

\hspace{-3mm}(b) \textit{Singular Matrix}: the matrix to be inverted is singular, i.e., $rcond<10^{-15}$,

\hspace{-3mm}(c) \textit{Number of Iterations Exceeded}: the number of iterations exceeds 10.

\begin{figure}
	\centering
	\includegraphics[width=0.99\linewidth]{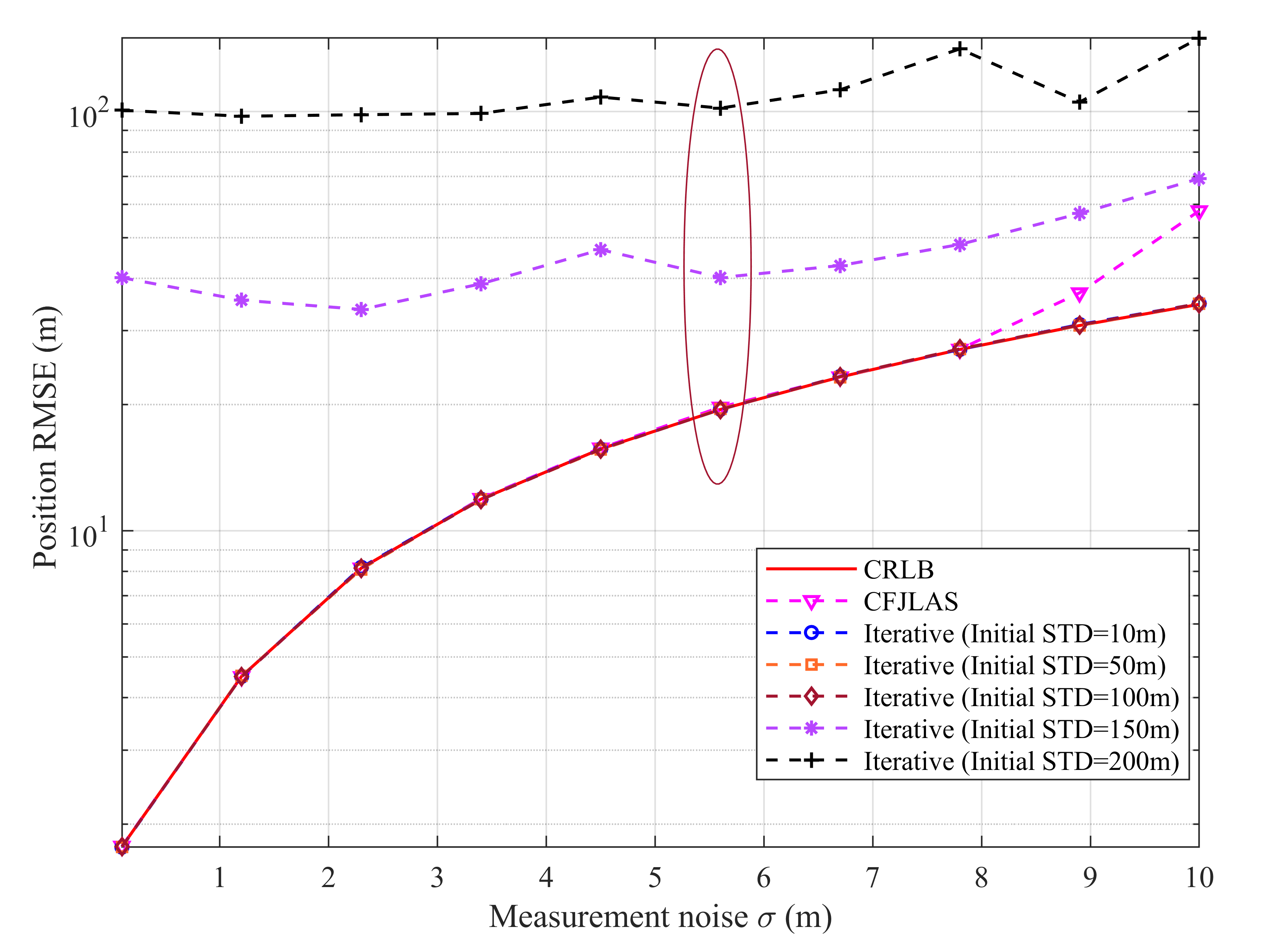}
	\caption{Position RMSE vs. measurement noise.  The position estimation error for the  iterative method is the average of the \textit{Convergence} case only. The iterative method is initialized with random position error of different STDs. At each measurement noise step, 100,000 simulations are run. The RMSEs of the iterative method initialized with large errors are much larger than that of the CFJLAS. More details on the results inside the brown ellipse are listed in Table \ref{table_Iterativecases}.}
	\label{fig:itererrorvsnoise}
\end{figure}

\begin{figure}
	\centering
	\includegraphics[width=0.99\linewidth]{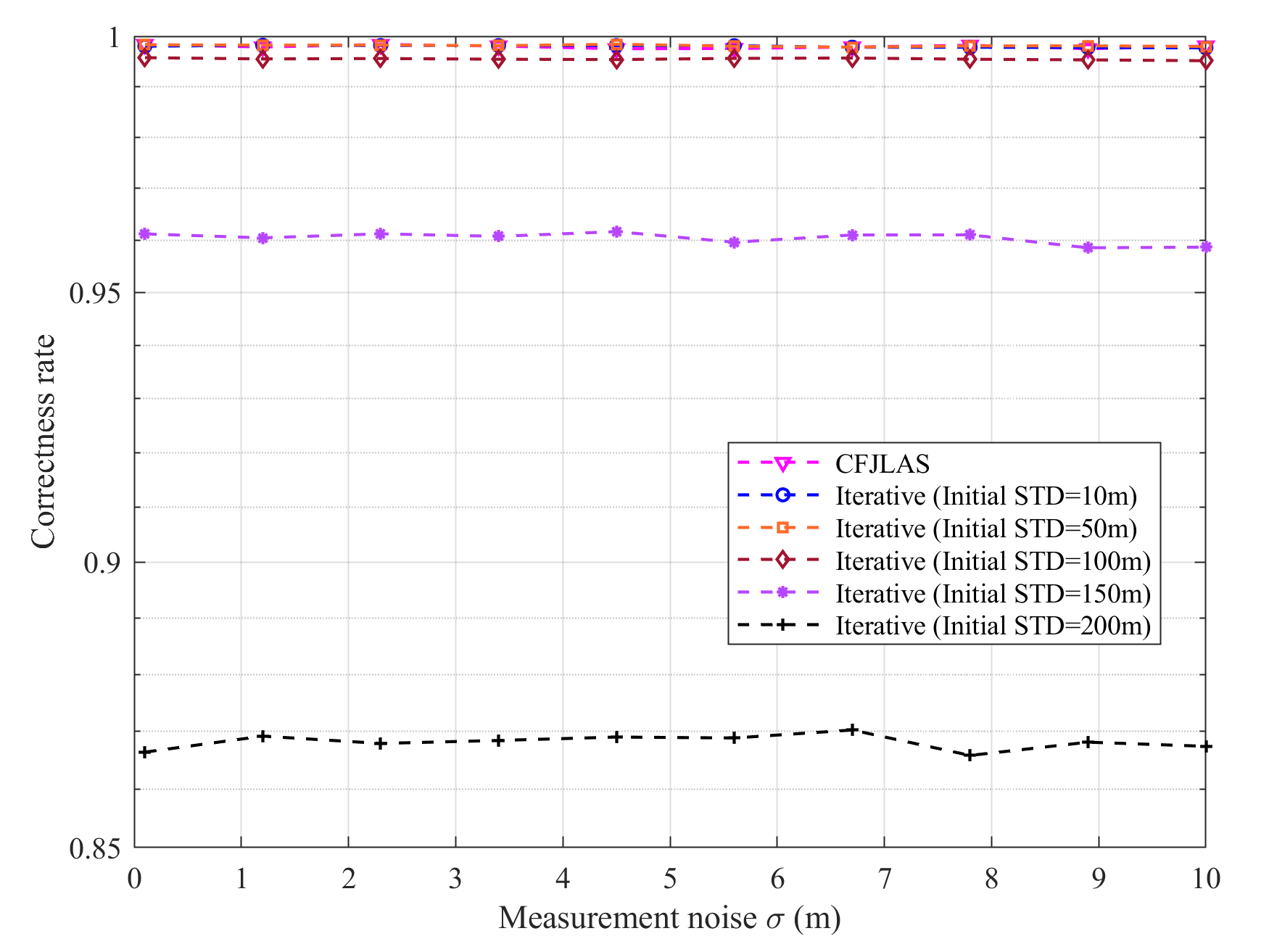}
	\caption{Correctness rate of CFJLAS and iterative method. The threshold to determine the correctness is set to $3\sqrt{\mathsf{CRLB}}$. The iterative method is initialized with different STDs of position error. 100,000 Monte-Carlo runs are done for each noise step. The proposed CFJLAS method has a correctness rate above 99.7\%, similar to that of the iterative method with initial STD=10 and 50 m. The correctness rate of the iterative method decreases significantly when the initial position error STD becomes larger than 100 m.}
	\label{fig:correctness}
\end{figure}

\hspace{-3.5mm}In addition, we use the same threshold $3\sqrt{\mathsf{CRLB}}$ to determine the correct result, denoted by ``(0) \textit{Correct Estimate}''. We first consider the \textit{Convergence} case (case (a)) only and illustrate the estimation errors in Fig. \ref{fig:itererrorvsnoise}. The probability of \textit{Correct Estimate} or correctness rate is shown in Fig. \ref{fig:correctness}.

It can be seen from Fig. \ref{fig:itererrorvsnoise} that the position estimation errors of the CFJLAS reach the CRLB when the measurement noise is small. In comparison, the iterative method shows a different performance depending on the initial position error. When the iterative method is initialized with small position errors (STD=10 and 50 m), the positioning results reach the CRLB. When the initial position errors are large (STD=150 and 200 m), the positioning results deviate from the CRLB even when the iterative method converges. The results of the iterative method with STD=100 m initialization error look the same as those with STD=10 and 50 m. This is because only the results from the \textit{Convergence} case are shown. And the actual performance when STD=100 m is inferior compared with when STD=10 and 50 m, as will be explained in the next two paragraphs.
	
As can be seen from Fig. \ref{fig:correctness}, the correctness rate of the CFJLAS method is close to 1 and is similar to that of the iterative method with the initial error STD=10 and 50 m. The correctness rate of the iterative method becomes smaller when STD=100 m and above, showing a performance degradation with larger initial position errors.

We take a close look at the results of the iterative method when measurement noise $\sigma=5.6$ m, which are within the brown ellipse in Fig. \ref{fig:itererrorvsnoise}. In the simulation runs for every situation with different number of ANs, we count the number of results that fall in the (a), (b) and (c) cases, and list them in Table \ref{table_Iterativecases}. As can be seen from the table, the numbers of results in the \textit{Correct Estimate} and the \textit{Convergence} cases both decrease when the initial position error grows. In addition, more results for STD=100 and above fall into (b) \textit{Singular Matrix} and (c) \textit{Number of Iterations Exceeded} cases. The table shows that the iterative method depends on proper initialization, and poor initial values will cause large errors in the final estimation results.

\begin{table*}[!t]
\centering
\begin{threeparttable}
	\caption{Number of Results for All Cases of the Iterative Method with Different Initial Error}
	\label{table_Iterativecases}
	\centering
	\begin{tabular}{c c r r r r r}
		\toprule
		\multicolumn{2}{c}{Initial position error STD (m)} & 10 & 50 & 100 & 150 & 200\\
		\hline
		{(0)}&{\textit{Correct Estimate} (error$<3\sqrt{\mathsf{CRLB}}$)}  &99,811&	99,801	&99,558	&95,958	&86,887	 \\
		(a) & \textit{Convergence} (increment norm$<0.01$ m) & 100,000	&100,000	&99,750	&96,096	&86,917	\\
		(b)& \textit{Singular Matrix} ($rcond<10^{-15}$) & 0	&0	&81	&1,329	&4,432	\\
		(c)& \textit{Number of Iterations Exceeded} ($\geq10$) & 0	&0	&169	&2,575	&8,651	\\
		\bottomrule
	\end{tabular}
	
	\begin{tablenotes}[para,flushleft]
	Note: The measurement noise $\sigma=$5.6 m. When the initial position error increases, the numbers of results in cases (0) and (a) reduce, and the numbers of results in cases (b) and (c) increase. This shows a dependence on the initialization error, i.e., a degradation in estimation performance with a growing initialization error. The numbers in cases (a), (b) and (c) have a sum of 100,000, showing that they cover all the cases for the iterative method.
	\end{tablenotes}
\end{threeparttable}
\end{table*}

\subsection{Computational Complexity}\label{compana}
To evaluate the computational complexity, we compare the new CFJLAS method and the iterative method. We notice that the iterative method requires more iterations or more computational complexity when the initial guess deviates from the true parameters. However, the complexity of the new CFJLAS is stable as analyzed in Section \ref{comana}.

We simulate the case with inaccurate initialization for the iterative method. We fix the measurement noise at $\sigma=2$ m, and vary the error STD of the initial position for the iterative method from 10 to 200 m. The 8-AN case is used. At each initial STD step, we run 100,000 Monte-Carlo simulations for each of the two methods.

\begin{figure}
	\centering
	\includegraphics[width=0.99\linewidth]{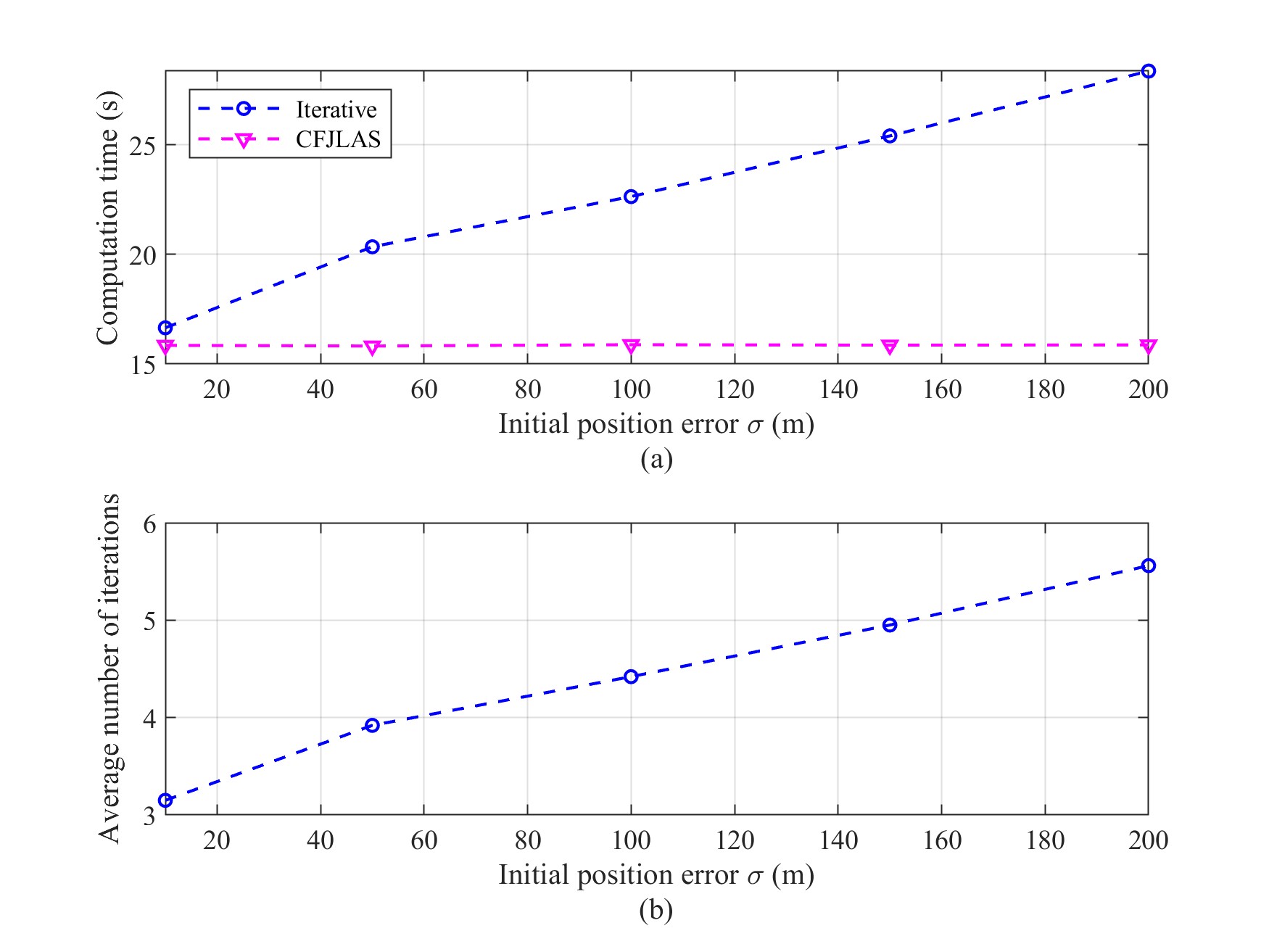}
	\caption{(a) Computation time (CFJLAS \& Iterative) vs. initial position error. (b) Average number of iterations (Iterative) vs. initial position error. The measurement noise $\sigma$ is fixed at 2 m. The computation time and the average number of iterations are the sum and average of 100,000 simulations, respectively. The number of iterations and computation time of the iterative method grow when the initial guess deviates from the true value. The computation time of the CFJLAS does not change with initial position error and is smaller than that of the iterative method. }
	\label{fig:complexityerr}
\end{figure}

The computation time of both methods are depicted in Fig. \ref{fig:complexityerr} (a). The average number of iterations for the iterative method from 100,000 Monte-Carlo runs is shown in Fig. \ref{fig:complexityerr} (b). We can see from the figure that the number of iterations increases when the initial guess has larger error. With inaccurate initialization, the computational complexity of the iterative method is growing and is larger than that of the CFJLAS. The result is consistent with the complexity analysis in Section \ref{compana}, and shows the superiority of the CFJLAS.

Nowadays, IoT devices with more functionalities and lower power consumption are developing fast. The reduced complexity and independence on initial guess make the new CFJLAS method more appealing to the power and resource constrained applications such as positioning for wearable devices, location-based service on smartphones, and autonomous guidance for miniature drones.

\section{Conclusion}
We propose a new closed-form method, namely CFJLAS, to solve the JLAS problem for the moving UNs using sequential one-way TOA measurements in a broadcast JLAS system. In this method, we transform the nonlinear relationship between the sequential TOA measurements and the UN's position, velocity, clock offset and clock skew parameters by squaring, differencing and employing two intermediate variables. Then, we reparameterize the problem with the two intermediate variables and convert it to two simultaneous quadratic equations, which can be solved analytically. After analytically obtaining the solution of the intermediate variables, an LS method is applied to roughly estimate the parameters of the moving UNs. Finally, a WLS step is applied to obtain the final optimal estimation.

We derive the CRLB of the estimation errors, analyze the theoretical error of the new CFJLAS method, and show that the new method reaches the CRLB under the small noise condition. We show that when the problem size is given, the complexity of the new CFJLAS is fixed while the conventional iterative method has a growing complexity with an increasing number of iterations. Compared with the iterative method, the new closed-form method does not require initial guess, obtains the solution in one shot, and thus has the advantage of low complexity.

Simulations in the 2D scene verify that the JLAS estimation accuracy of the new CFJLAS reaches the CRLB. Results also show that the new method does not require initialization and can obtain a high-accuracy estimation under the small noise condition, compared with the iterative method, which needs a proper initialization to obtain the correct result. Moreover, unlike the iterative method whose complexity increases with the number of iterations, the proposed CFJLAS is non-iterative and has a low computational complexity.

\appendices

\section{Solution to Quadratic Equations (\ref{eq:quadratic1}) and (\ref{eq:quadratic2})}
\label{Appendix1}
The solution to a pair of quadratic equations such as (\ref{eq:quadratic1}) and (\ref{eq:quadratic2}) can be found in \cite{zhao2020closed}. For the convenience of the interested readers, we put the method here in this appendix. Note that the notations used herein are only effective within this appendix.

We replace the unknowns with $x$ and $y$, respectively, and rewrite the two quadratic equations of (\ref{eq:quadratic1}) and (\ref{eq:quadratic2}) as 
\begin{equation} \label{eq:A001quadratic1}
	\begin{split}
		a_1 x^2+b_1 x y +c_1 y^2 + d_1 x+e_1 y+f_1=0 \text{,}
	\end{split}
\end{equation}
\begin{equation} \label{eq:A002quadratic2}
	\begin{split}
		a_2 x^2+b_2 xy +c_2 y^2 + d_2 x+e_2 y+f_2=0 \text{.}
	\end{split}
\end{equation}

We first remove the term $y^2$ by multiplying $c_1$ and $c_2$ to (\ref{eq:A002quadratic2}) and (\ref{eq:A001quadratic1}), respectively, and then subtracting the resulting equations. After re-organizing, we obtain
\begin{equation} \label{eq:A004yequation}
	(t_{1}x+t_2)y=t_3 x^2+t_4 x+t_5 \text{,}
\end{equation}
where
$$
t_1=b_1 c_2-b_2 c_1 \text{,} \; t_2=e_1 c_2 -e_2 c_1 \text{,}
$$
$$
t_3=-a_1 c_2+ a_2 c_1, \; t_4 = -d_1 c_2 + d_2 c_1 \text{,}
$$
$$
t_5=-f_1 c_2 + f_2 c_1 \text{.}
$$

Here are two cases. One is $t_{1}x+t_2=0 $ and the other is $t_{1}x+t_2 \neq 0 $.

Case 1: $t_{1}x+t_2=0 $

If $t_1=0$, then $t_2$ must equal to zero. In this sub-case, the problem reduces to solving the equation of 
\begin{equation} \label{eq:A005xequation}
	t_3 x^2+t_4 x+t_5=0 \text{.}
\end{equation}

After substituting the root of $x$ from (\ref{eq:A005xequation}) into (\ref{eq:A002quadratic2}), the root of $y$ can be found.

If $t_1\neq 0$, then we need to test if $-t_2/t_1$ is the root of $x$ by substituting it into (\ref{eq:A005xequation}). If it satisfies (\ref{eq:A005xequation}), then the root of $y$ can be found by substituting $x$ into (\ref{eq:A002quadratic2}). Otherwise, there is no solution.

Case 2: $t_{1}x+t_2\neq 0 $

In this case, we have 
\begin{equation} \label{eq:A007commoncase}
	y=(t_3 x^2+t_4 x+t_5)/(t_{1}x+t_2) \text{.}
\end{equation}

By substituting (\ref{eq:A007commoncase}) into (\ref{eq:A002quadratic2}), we come to a quartic equation in $x$ as
\begin{equation} \label{eq:A009quartic}
	\alpha x^4+\beta x^3+ \gamma x^2 +\lambda x +\mu=0 \text{,}
\end{equation}
where
$$
\alpha=a_1 t_1^2 + b_1 t_1 t_3 + c_1 t_3^2 \text{,}
$$
$$
\beta=d_1 t_1^2 + 2 a_1 t_1 t_2 + b_1 t_1 t_4 + b_1 t_2 t_3 + 2 c_1 t_3 t_4 + e_1 t_1 t_3 \text{,}
$$
$$
\begin{array}{rr}
	\gamma =c_1 (t_4^2 + 2 t_3 t_5) + a_1 t_2^2 + f_1 t_1^2 + b_1 t_1 t_5\\
	+ b_1 t_2 t_4 + 2 d_1 t_1 t_2 + e_1 t_1 t_4 + e_1 t_2 t_3 \text{,}
\end{array} 
$$
$$
\lambda=d_1 t_2^2 + b_1 t_2 t_5 + 2 c_1 t_4 t_5 + e_1 t_1 t_5 + e_1 t_2 t_4 + 2 f_1 t_1 t_2 \text{,}
$$
$$
\mu=f_1 t_2^2 + e_1 t_2 t_5 + c_1 t_5^2 \text{.}
$$

Solution of the quartic equation (\ref{eq:A009quartic}) can be found in the mathematical literature such as \cite{shmakov2011universal,riccardi2009solution}. We simply write the solution as follows in this appendix without derivation for those interested readers.

There are at most four roots for this equation, either real or complex values. The general form of the roots is given by
\begin{equation} \label{eq:A010xroots}
	\begin{split}
		x(1), x(2)=-\frac{\beta}{4\alpha} - s \pm \frac{1}{2}\sqrt{-4s^2-2p+\frac{q_0}{s}} \text{,}\\
		x(3), x(4)=-\frac{\beta}{4\alpha} + s \pm \frac{1}{2}\sqrt{-4s^2-2p-\frac{q_0}{s}} \text{,}
	\end{split}
\end{equation}
with the variables expressed as follows,
\begin{align}\label{eq:vareq}
s=\frac{1}{2} \sqrt{-\frac{2}{3}p+\frac{1}{3\alpha} (q_1+\frac{\Delta_0}{q_1})}\text{,}
\;p=\frac{8 \alpha \gamma - 3 \beta^2}{8 \alpha^2} \text{,}\nonumber\\
q_0=\frac{\beta^3 - 4 \alpha \beta \gamma + 8 \alpha^2 \lambda}{8 \alpha^3}\text{,} \;q_1=\sqrt[3]{\frac{\Delta_1+\sqrt{-27\Delta}}{2}} \text{,}\nonumber\\
\Delta = -\frac{\Delta_1^2-4\Delta_0^3}{27}\text{,}\;\Delta_0=\gamma^2-3 \beta \lambda +12 \alpha \mu \text{,}\nonumber\\
\Delta_1=2 \gamma^3 -9 \beta \gamma \lambda +27 \beta^2 \mu+27 \alpha \lambda^2 -72\alpha \gamma \mu \text{.}
\end{align}

\section{Complexity Analysis for CFJLAS and Iterative Method}
\label{Appendix2}

We assume that a single addition, multiplication or square root operation for real numbers can be done in one flop, following \cite{golub2013matrix}. By counting the flops of the major operations in the CFJLAS and the iterative method, their complexities can be estimated.

\subsection{CFJLAS Method}
For the CFJLAS method, we first investigate the complexity of the raw estimate \eqref{eq:LSestimate} and the WLS compensation (\ref{eq:refine1}). Following \cite{quintana2001note}, we use $2k^3$, where $k$ is the number of rows (or columns) of a square matrix, to approximate the number of flops required for the matrix inverse operation. The number of flops required by \eqref{eq:LSestimate} is
\begin{align} \label{eq:compLS}
&2(2K+2)^2(M-1)+2(2K+2)^3+\nonumber\\
&2(2K+2)^2+2(2K+2)(M-1)+5M-5\nonumber\\
=&
16K^3 +8K^2 M + 48K^2 + 20KM + 48K + 17M + 15,
\end{align}
where $K$ is the dimension of the position and $M$ is the number of ANs.

Considering that $\bm{W}$ is a diagonal matrix, the number of flops required by (\ref{eq:refine1}) is
\begin{align} \label{eq:comWLS}
&2(2K+2)^2M+2(2K+2)^3+2(2K+2)M+\nonumber\\
&2(2K+2)^2+(2K+2)M+2M\nonumber\\
=&16K^3 + 8K^2M+56K^2 + 22KM+64K + 16M + 24.
\end{align}

We then look into the complexity of Step 2: Solution for Intermediate Variables given by Section \ref{solutionIV}. The coefficients of the equations \eqref{eq:quadratic1} and \eqref{eq:quadratic2} are obtained by the equations below them. We take $a_1$ as an example. By observing $\bm{H}_1$, we can see that there are $K+1$ entries with the value of 1. Therefore, $a_1=[\bm{U}]_{:,1}^T\bm{H}_1[\bm{U}]_{:,1}$ takes $(K+1)$ multiplications and $K$ additions, and thus $(2K+1)$ flops in total. All the six coefficients with the subscript of ``1'' are similar and they take $6(2K+1)$ flops in total for $a_1\sim f_1$. We also notice that $\bm{H}_2$ has $(2K+2)$ entries of one and all other entries of zero in it. Thus, the total number of flops for computing $a_2\sim f_2$ is $6(4K+3)$. Hence, we have 
\begin{align}\label{eq:complexcoef1}
6(6K+4)
\end{align}
flops for computing the coefficients of the two equations.

We then count the flops of computing the coefficients in \eqref{eq:A004yequation}, and obtain $15$. The number of flops required by computing the coefficients in \eqref{eq:A009quartic} is $81$. For finding the roots of $x$ in \eqref{eq:A010xroots}, there may be operations of complex numbers. We consider there are 2 flops for a complex addition and 6 flops for a complex multiplication or square root. Therefore, solution for the roots of $x$ in \eqref{eq:A010xroots} takes $4\times 38 =152$ flops. Computing the variables in \eqref{eq:vareq} requires $278$ flops. Then, computing the roots of $y$ in \eqref{eq:A007commoncase} requires $4 \times 28 =112$ flops. Hence, finding the roots of the quadratic equations \eqref{eq:quadratic1} and \eqref{eq:quadratic2} needs
\begin{align} \label{eq:complexroots}
15+81+152+278+112=638
\end{align}
flops.

Root selection given by \eqref{eq:residual} needs to compute the residual as given by \eqref{eq:residual1}. Computing $\tilde{\boldsymbol{p}}+\tilde{\boldsymbol{v}}t_i-\hat{\boldsymbol{p}}_i$ needs $K$ multiplications and $2K$ additions. The norm operation needs $K$ multiplications, $(K-1)$ additions and 1 square root. Thus, there are 5$K$ flops for each measurement. We need 3 additions and 1 multiplications to compute one residual $[\boldsymbol{r}]_{i}$. Considering $M$ measurements and at most 4 roots from the two equations, we need
\begin{align}\label{eq:complexsel}
	4M(5K+4)
\end{align}
flops to compute all the residuals. The root selection in \eqref{eq:residual} has $2M$ multiplications and $(2M-1)$ additions. For 4 roots, there are
\begin{align}\label{eq:complexsel2}
	4(4M-1)
\end{align}
flops in total.

As a result, we obtain the total number of flops by summing \eqref{eq:compLS} to \eqref{eq:complexsel2} as
\begin{align} \label{eq:complexDapp}
D \approx& 32K^3 +16K^2M + 104K^2+ 62KM +\nonumber\\
 &148K + 65M + 697
\end{align}

\subsection{Iterative Method}
The complexity in a single iteration, denoted by $L$, is mainly comprised of matrix multiplication and matrix inversion. Following the iterative method that estimates position, velocity, clock offset and skew in \cite{zhao2020optloc}, the matrix multiplication and matrix inverse operations are similar to that of (\ref{eq:refine1}). Therefore, it has the same expression as \eqref{eq:comWLS}, i.e.,
\begin{align}
L\approx16K^3 + 8K^2M+56K^2 + 22KM+64K + 16M + 24.
\end{align}

In most of the time, the iterative method requires more than one iteration to obtain the correct result. If there are $n$ iterations, then the total complexity is $nL$.


%


\ifCLASSOPTIONcaptionsoff
    \newpage
\fi



\bibliographystyle{IEEEtran}
\bibliography{IEEEabrv,paper}
\end{document}